\documentclass[12pt]{iopart}
\usepackage{graphicx}

\usepackage{iopams}

\usepackage{amsfonts}   

\begin{document}

\title[Numerical implementation of DMFT: application to the rLV model]{Numerical implementation of dynamical mean field theory for disordered systems: application to the Lotka-Volterra model of ecosystems}

\author{F Roy$^1$, G Biroli$^{2}$, 
G Bunin$^3$ and C Cammarota$^4$}

\address{$^1$ Institut de physique th\'{e}orique, Universit\'{e} Paris Saclay,
CEA, CNRS, F-91191 Gif-sur-Yvette, France}
\address{$^2$ Laboratoire de Physique de l'Ecole Normale Superieure, ENS, Universit\'{e} PSL, CNRS, Sorbonne Universit\'{e}, Universit\'{e} Paris-Diderot, Sorbonne Paris Cit\'{e}, Paris, France}
\address{$^3$ Department of Physics,
Technion-Israel Institute of Technology, Haifa 32000, Israel}
\address{$^4$ King's College London, Department of Mathematics, Strand, London WC2R 2LS, United Kingdom}
\ead{felix.roy@ipht.fr}

\begin{indented}
\item[]January 2019
\item[]{\it Keywords}: Disordered systems, Non-equilibrium dynamics, Population dynamics.
\item[] \submitto{\jpa}
\end{indented}

\begin{abstract}
Dynamical mean field theory (DMFT) is a tool that allows to analyze the stochastic dynamics of $N$ interacting degrees of freedom in terms of a self-consistent $1$-body problem. In this work, focusing on models of ecosystems, we present the derivation of DMFT through the dynamical cavity method, and we develop a method for solving it numerically. Our numerical procedure can be applied to a large variety of systems for which DMFT holds. We implement and test it for the generalized random Lotka-Volterra model, and show that complex dynamical regimes characterized by chaos and aging can be captured and studied by this framework. 
\end{abstract}

%

\maketitle
\vspace{10pt}

%
%
%

\section{Introduction}
\label{sec-introduction}
A growing body of work has demonstrated that the properties of communities of interacting species can be studied using tools of statistical mechanics, with the role of the thermodynamic limit being played by the large number of species. Such high-diversity communities, with tens to thousands of species, are ubiquitous and can be found anywhere from microbes in the gut to plants in a rain forest \cite{faust_microbial_2012}. Most of the works have focused on the properties of fixed-points of the dynamics, and much less is known about the dynamics themselves, in particular when they never reach a fixed-point. \\
Dynamical mean-field theory (DMFT) is a useful theoretical framework which has been often used in the past to study complex stochastic dynamics of interacting degrees of freedom (spins, agents, neurons, ...) \cite{sompoZip_dmft, cugliandolo1993analytical, galla_replicator, opperPRL,sompoPRL}. In this work we develop DMFT for models 
of ecosystems formed by a large number of interacting species \cite{biroli_marginal, buninPRE, galla_rLV}. 
In the limit of large ecosystems, interactions between different species are commonly modeled by taking random interaction strengths \cite{may_theoretical_2007,barbier_assembly}. The resulting model consists in generalized Lotka-Volterra equations with random couplings. This leads to interesting problems of statistical physics, similar to ones encountered in the theory of disordered systems. Yet, there are a number of crucial differences; in particular an ecosystem is driven by non-conservative forces, hence its dynamics cannot be mapped in general to the one of a physical system in thermal equilibrium. This leads to complex dynamical regimes which have been discussed in other fields before, mainly in neural networks \cite{sompoPRL} and game theory \cite{galla_replicator, opperPRL} (see also \cite{berthier2000two}).\\
Solving numerically the equations corresponding to DMFT represents a major difficulty due to the retarded friction kernel 
and the non-linearity of the equations. In the past this obstacle
has been solved---actually circumvented---only for simplified (spherical or truncated) spin-glass models
for which DMFT equations greatly simplify and reduce to closed integro-differential equations on correlation and response functions of local degrees of freedom \cite{kim2001dynamics}. 
To the best of our knowledge, a procedure to numerically integrate DMFT is still missing (with the exception of 
\cite{eissfeller_numDMFT} that was restricted to the case of stationary dynamics and Ising spins), especially 
one able to analyze the complex dynamics relevant for ecosystems.  
In this work, following ideas developed for DMFT of strongly correlated quantum systems \cite{georges_dmft}, we develop a generic numerical scheme to solve DMFT. Our method lays foundations for the study  of high-diversity ecological dynamics, but 
also provides general tools that can be applied to problems beyond ecology, for example in 
the fields mentioned above \footnote{Our code is available in a public gitHub repository \cite{felixroy_implements_2019}.}.

We focus on the generalized Lotka-Volterra model of ecosystems.
We first present a derivation of DMFT based on the dynamical cavity method \cite{parisi_spin}, which is a more intuitive procedure compared to the usual ones based on generating functional formalism, such as Martin-Siggia-Rose-DeDominicis-Janssen \cite{galla_replicator, galla_rLV, eissfeller_numDMFT}. 
We then detail our numerical approach for solving DMFT, and show concrete examples of its implementation. 
This allows us to test the method, and illustrate its ability to describe and characterize complex dynamics involving chaos and aging. 
We finally conclude by discussing further directions and possible future applications.

\section{The random Lotka-Volterra model}

In this section, we introduce the random Lotka-Volterra (rLV) model \cite{buninPRE}, which describes the dynamics of interacting species. We present its phase portrait in the limit of a large number of species.

\subsection{Definition and notations}
The ecosystem consists of $S$ species. Each species $i$ is characterized by its population $N_i(t)$ which is 
a positive continuous variable at all times $t$. In the absence of interactions each species may grow until saturation (e.g., due to limitations on resources).
The impact of other species is modeled through bilinear interactions. A small immigration rate $\lambda_i$ is added so that new individuals arrive to the ecosystem from the outside. The dynamical equations read:

\begin{equation*}
\forall i=1..S, \quad \frac{d N_i}{dt}=\frac{r_i}{K_i} N_i (K_i - N_i) - N_i \sum_{j \neq i} \alpha_{ij} N_j + \lambda_i
\end{equation*}

The different parameters are the intrinsic growth rates $r_i$ of the species, their single-species population size (carrying capacities) $K_i$ in the environment and the interaction matrix $\alpha$. Within our convention, a positive coefficient $\alpha_{ij}$ indicates that the presence of species $j$ is deleterious to the species $i$, due to predation or competition over resources.

For clarity of presentation, in this article we mainly discuss the case where all $r_i$ and $K_i$ are set to unity, and the immigration rate $\lambda$ is uniform. The analytical and numerical tools described can be used more generally. Immigration will act as a regularization of the problem. All results will be derived with infinitesimal but finite immigration rate $\lambda>0$. We will separately discuss in the last section the case without immigration. The elements of the interaction matrix $\alpha_{ij}$ are i.i.d. Gaussian random variables with moments:
$$\overline{\alpha_{ij}}=\mu/S, \qquad \overline{(\alpha_{ij}-\overline{\alpha_{ij}})^2}=\sigma^2/S, \qquad \overline{(\alpha_{ij}-\overline{\alpha_{ij}})(\alpha_{ji}-\overline{\alpha_{ji}})}=\gamma\sigma^2/S $$

The scaling with $S$ ensures a proper large $S$ limit. In this limit, the model becomes characterized by three parameters only: the average strength of interaction $\mu$, the variety of interactions $\sigma$, and their symmetry $\gamma$. More specifically, $\gamma$ ranges from -1 (fully antisymmetric case, where all interactions are of predation-prey type) to 1 (fully symmetric case, where an energy can be defined).

For a real ecosystem with given size $S$, the parameters $\mu$, $\sigma$ and $\gamma$ can be statistically computed from the interaction matrix $\alpha$. Our result then stands for this ecosystem with the relevant parameters value. From numerical simulations, we find that ecosystems with $S>200$ 
are well described by results obtained in the "thermodynamic" limit $S\rightarrow \infty$.

\subsection{Phase diagram}

\begin{figure}
\centering
\begin{minipage}{.7\textwidth}
  \centering
  \includegraphics[width=\linewidth]{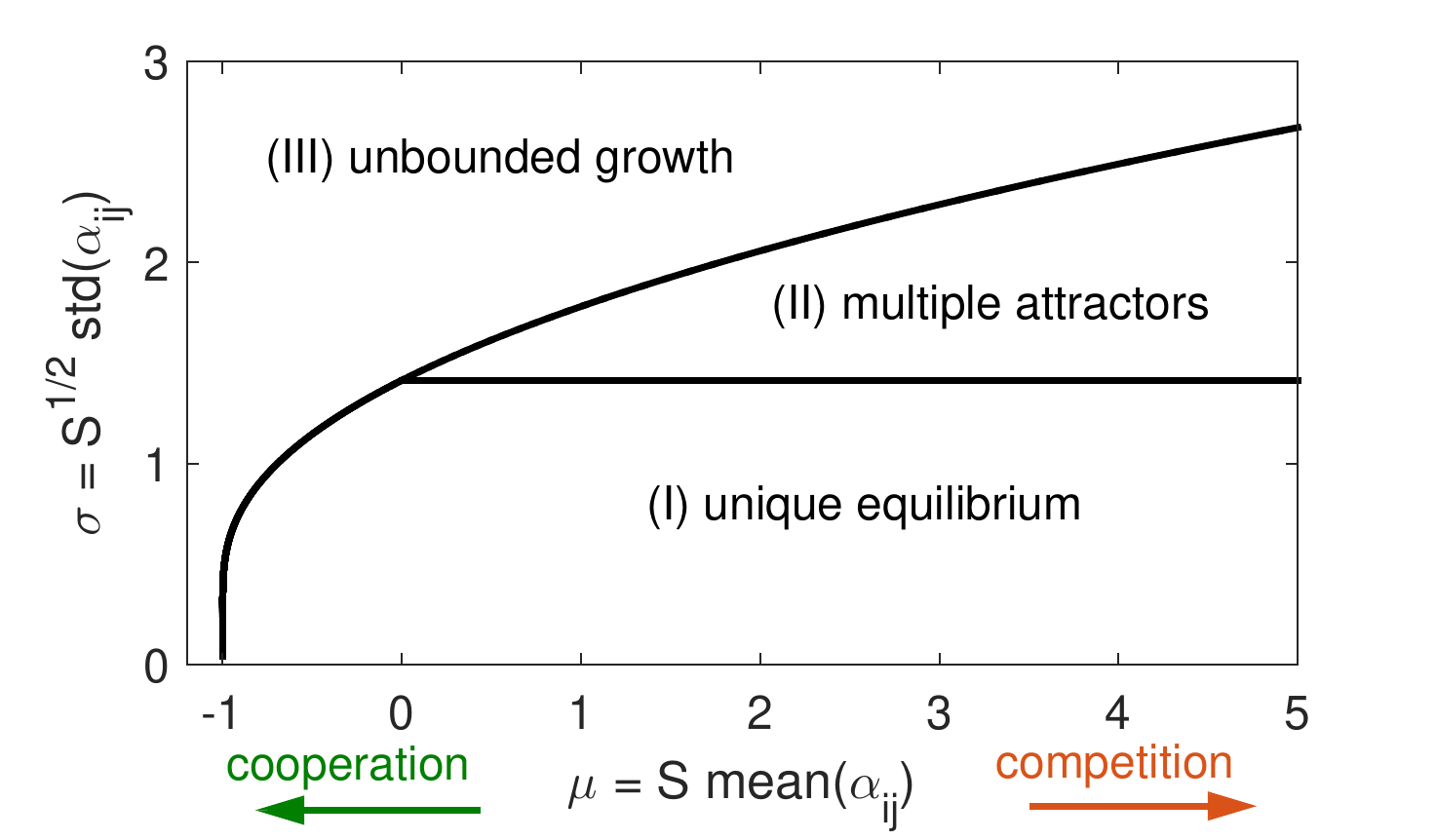}
\end{minipage}%
\begin{minipage}{.3\textwidth}
  \centering
  \includegraphics[width=\linewidth]{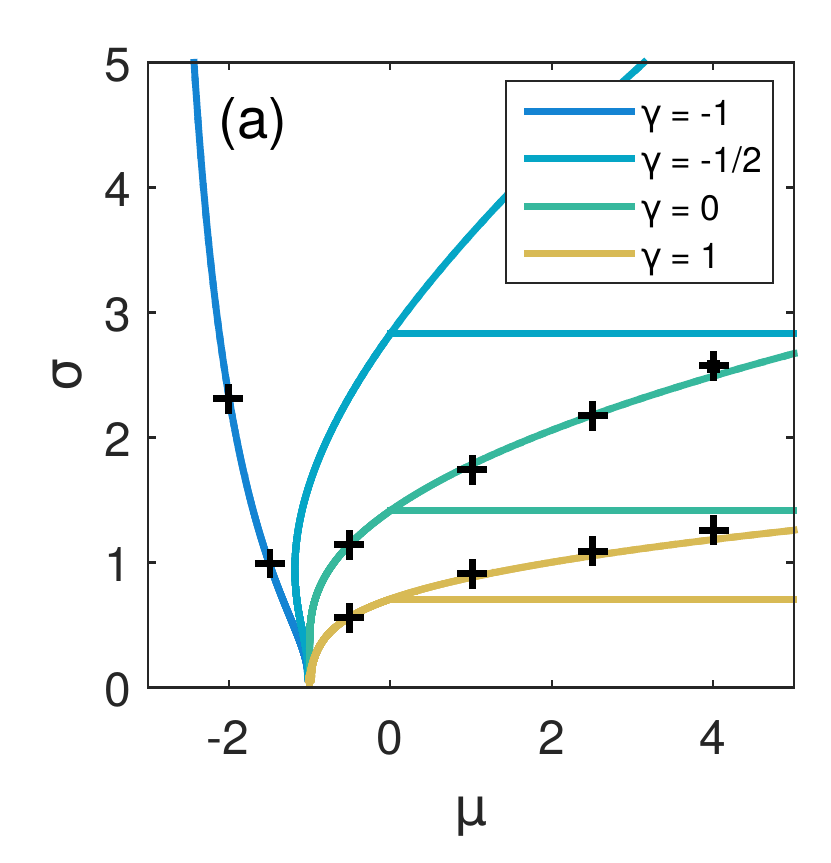}
\end{minipage}
\caption{ Phase diagram, taken from \cite{buninPRE}. Left: At $\gamma=0$, the transition from Unique Equilibrium to Multiple Attractors is independent from $\mu$ and lies on the line $\sigma=\sqrt{2}$. Right: For generic $\gamma$ the transition lies on the line $\sigma=\frac{\sqrt{2}}{1+\gamma}$. Increasing the symmetry $\gamma$ shifts the transitions towards lower variance and stronger interactions. This result shows how predation-prey relations may stabilize an ecosystem.} 
\label{fig:phaseP}
\end{figure}

In the large-$S$ limit, three different dynamical phases are found \cite{buninPRE}; see \fref{fig:phaseP}.

\begin{itemize}

\item {\it Phase I: Unique Equilibrium}. 
In this regime, corresponding to small $\sigma$, 
the ecosystem displays only one stable equilibrium. Whatever the initial conditions, each species asymptotically ends up with a given number of individuals which is always the same (it can be zero as some species go extinct). This equilibrium state is stable to local and global perturbations. On \fref{fig:1eq_traj} we display the dynamics of an ecosystem in this phase: each line represents the time evolution of the population of one species.

\begin{figure}
\centering
\begin{minipage}{.5\textwidth}
  \centering
  \includegraphics[width=\linewidth]{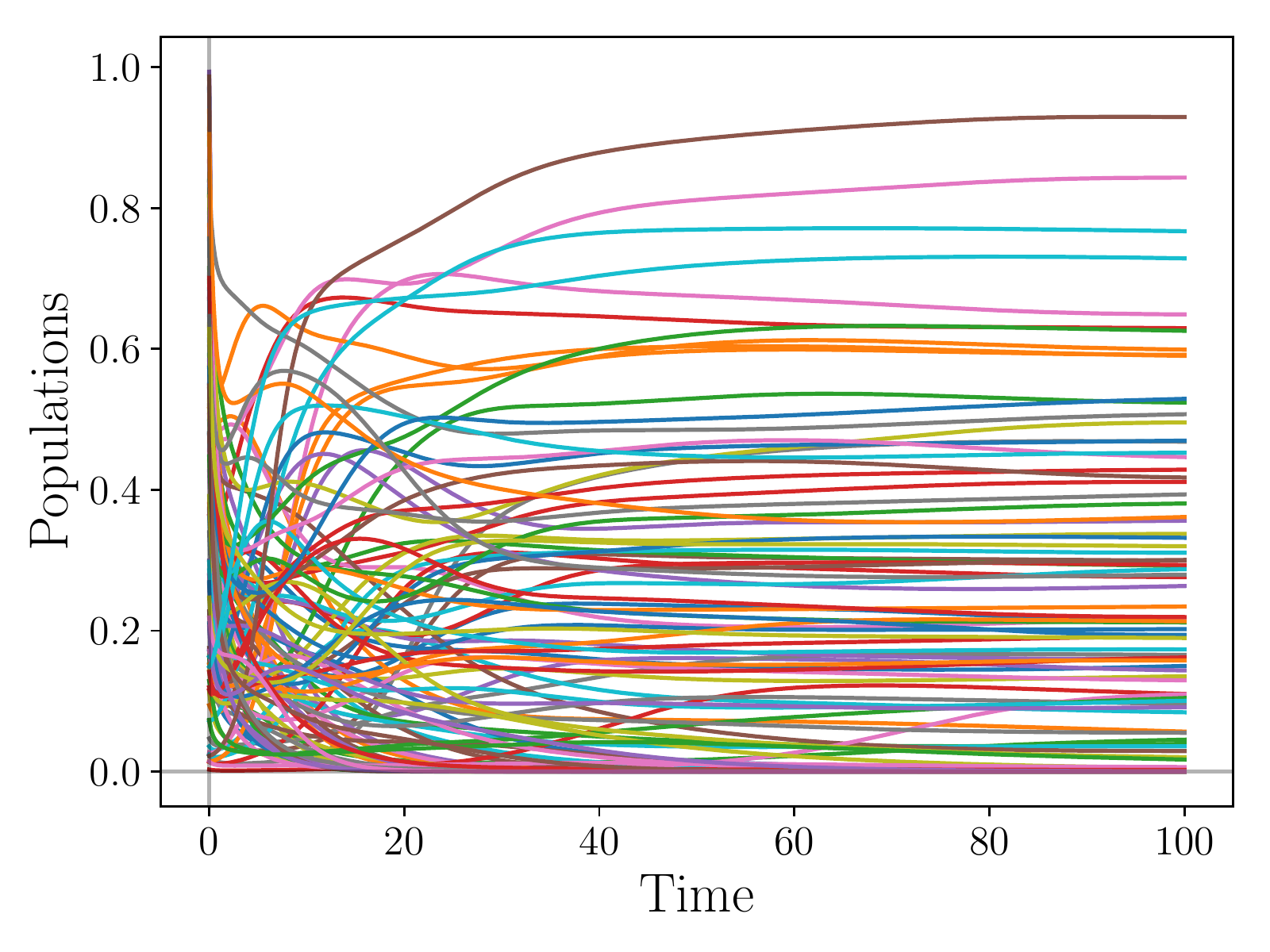}
\end{minipage}%
\begin{minipage}{.5\textwidth}
  \centering
  \includegraphics[width=\linewidth]{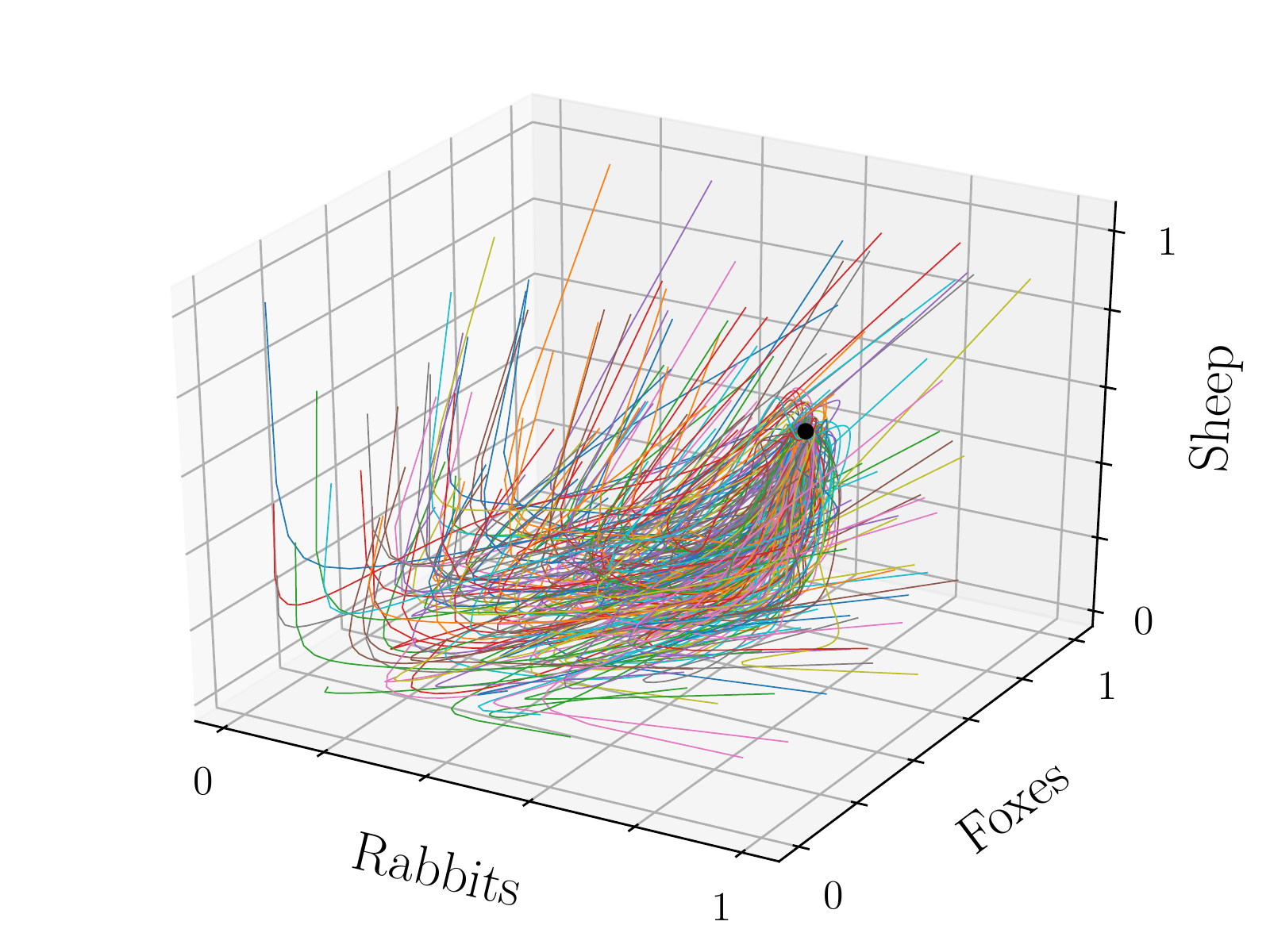}
\end{minipage}
\caption{Time evolution of 100 species in the Unique Equilibrium phase: 
$(\mu,\sigma,\gamma, \lambda|S)=(4,1,0,10^{-10}|100)$. 
Left: 
After a transient time, each species reaches a final population value which is stable. 
Right: 
Dynamical evolution of three species (e.g., `sheep', `rabbits' and `foxes') out of all the species. We show different trajectories obtained starting from different initial conditions. They always converge to the same equilibrium value (black dot) independent of the initial conditions, demonstrating the stability and uniqueness of this equilibrium.}
\label{fig:1eq_traj}
\end{figure}

\item {\it Phase II: Multiple Attractors}. When the variability in the interactions $\sigma$ is increased, the single stable fixed point loses its stability, 
and the system is left with a huge number of (possibly unstable) equilibria. This phase exhibits a complex dynamics with chaos (or aging dynamics for $\gamma=1$). An example of such dynamics can be seen on \fref{fig:chaosUG_traj}.

\item {\it Phase III: Unbounded  Growth}. When the average interaction is negative enough ($\mu < -1$), the interactions are cooperative enough to have a beneficial effect on any given species that overrides the single-species saturation. 
If we fix a higher $\mu$ and increase the standard deviation $\sigma$, at some point a small community of species will have cooperative interactions stronger than their own saturation and this subgroup of species will thus grow without bound, even though all the other species will die out. 
This explains the existence of phase III also for $\mu>-1$ for a large-enough  $\sigma$. An example of such dynamics is displayed in \fref{fig:chaosUG_traj}. It should be noted that the divergence occurs as a finite time explosion of the ecosystem. 
The unbounded growth is a pathology of the model that could be cured by a saturation stronger than quadratic. 

\begin{figure}
\centering
\begin{minipage}{.5\textwidth}
  \centering
  \includegraphics[width=\linewidth]{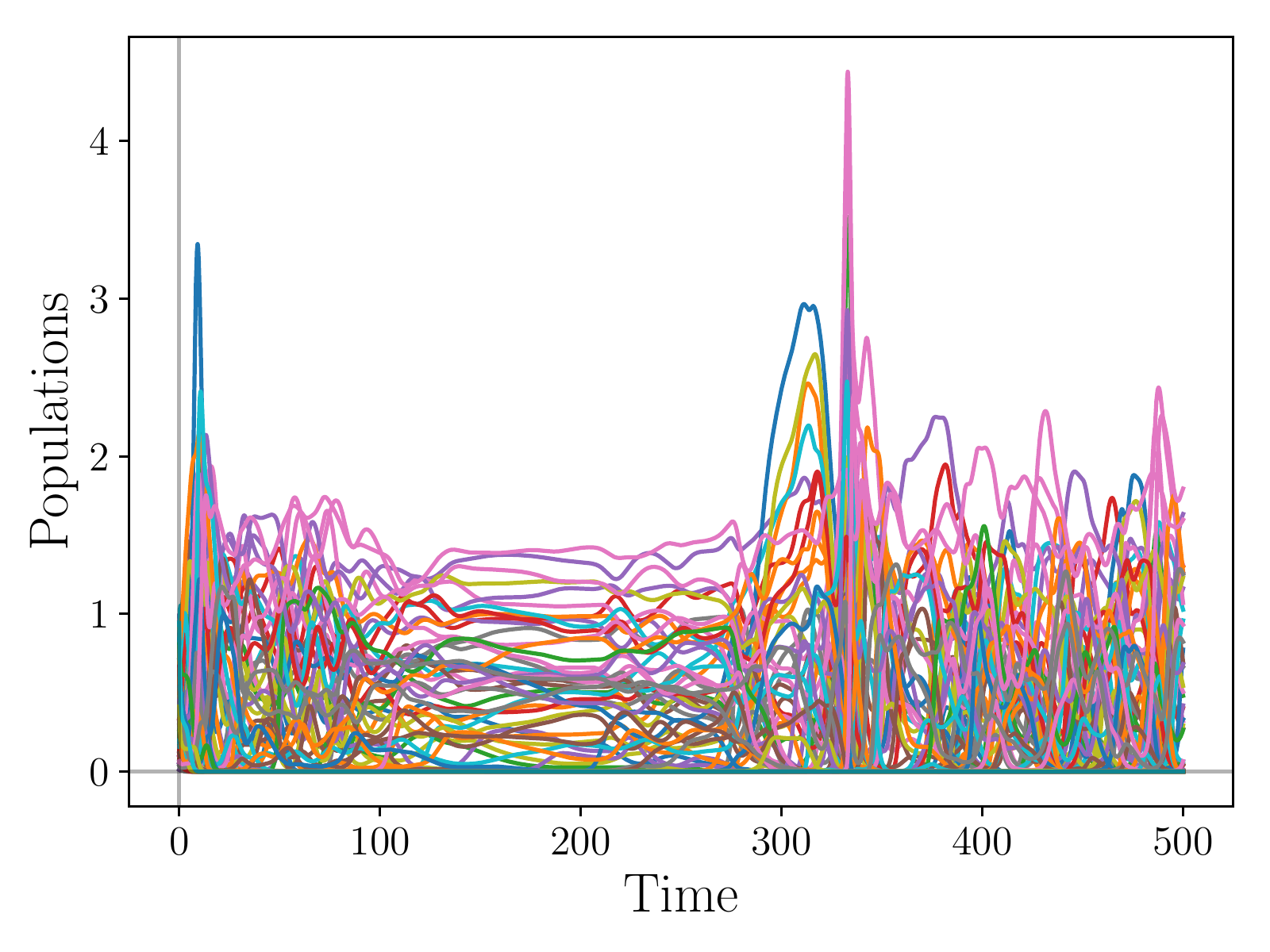}
  \label{fig:test1}
\end{minipage}%
\begin{minipage}{.5\textwidth}
  \centering
  \includegraphics[width=\linewidth]{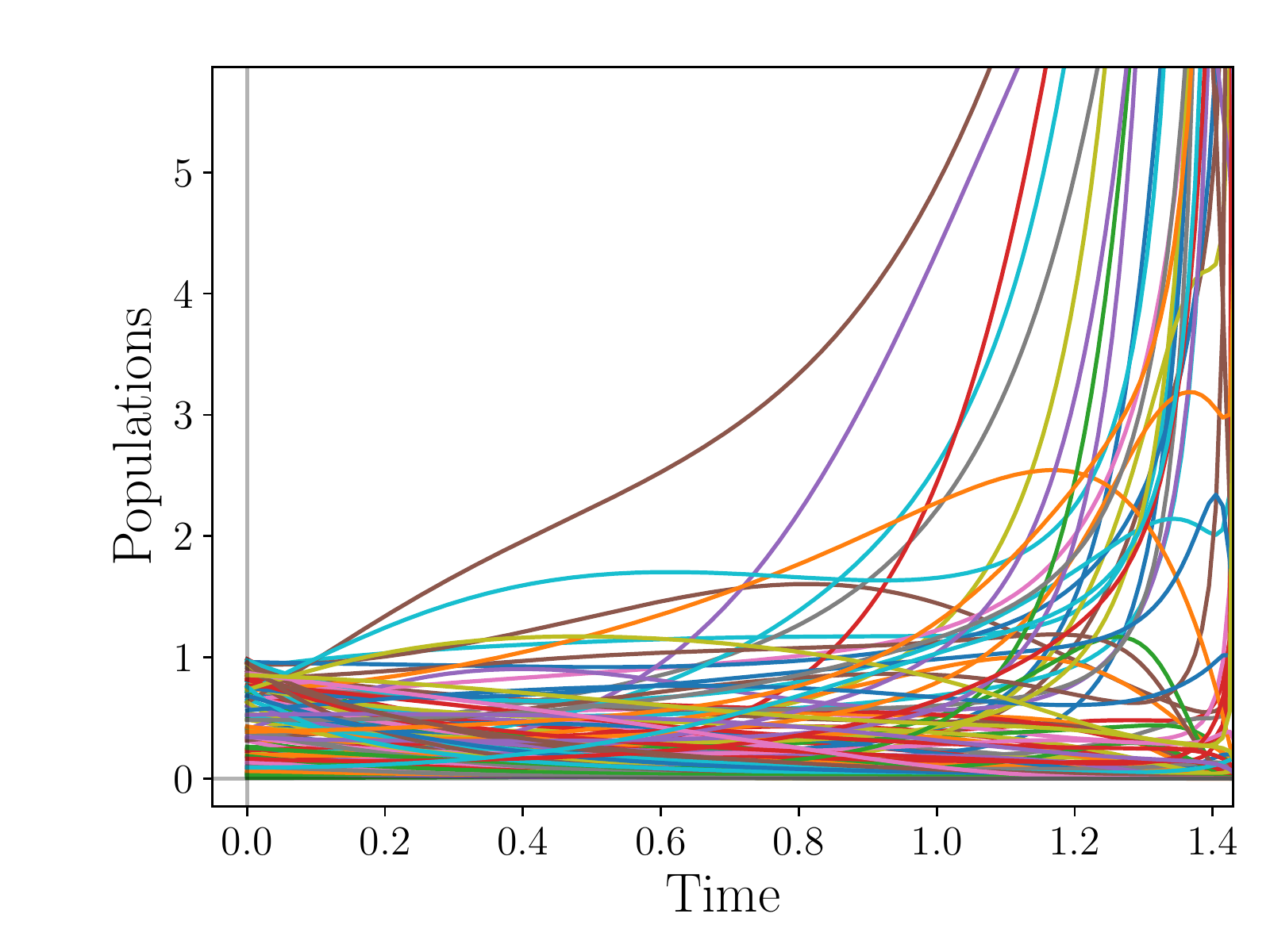}
  \label{fig:test2}
\end{minipage}
\caption{Left: Time evolution of 100 species in the Multiple Attractors phase: $(\mu,\sigma,\gamma,\lambda | S) =(4,2,0,10^{-10}|100)$. The trajectories do not display any simple behaviour; at some points, the system seems to relax to a fixed point before realizing that it has some unstable directions, and the dynamics starts again. Right: Time evolution of 100 species in the Unbounded Growth phase: $(\mu,\sigma,\gamma,\lambda | S) =(4,4,0,10^{-10}|100)$. A large proportion of species present a divergence of their population, while the other ones die out.}
\label{fig:chaosUG_traj}
\end{figure}

\end{itemize}
The borders between phases can be computed analytically: I/II and I/III are exact, but II/III is only approximate \cite{buninPRE}. They are shown on \fref{fig:phaseP}.

The symmetric case $\gamma=1$ is special in the sense that the Multiple Attractor phase is not a chaotic one, but rather a spin glass one \cite{biroli_marginal}: the dynamics gets slower and slower as the system approaches marginally stable states.  In this case, a kind of physical energy can be defined and serves as a Lyapunov function. The dynamics corresponds to a gradient descent in a rough energy landscape.

\section{The Dynamical Mean Field Theory}

In this section, we derive the Dynamical Mean Field Theory (DMFT) using the dynamical cavity method \cite{parisi_spin}. For simplicity, we first present the derivation in the simplest case of random Lotka-Volterra model, then we extend the result to more general models, and finally we explain a numerical method to solve the DMFT equation. We also checked the relevance of the description by comparing the DMFT results with direct simulations, increasing the size $S$ of the ecosystem.

\subsection{Derivation via the cavity method}
For simplicity, DMFT is first derived with the simplest random Lotka-Volterra model, presented in equation \eref{eq:rLV}. Our approach holds in more general cases, we will present its generalization in \sref{sec:DMFTGen}. We start from the Lotka-Volterra equations:

\begin{equation}
\forall i =1..S, \quad \quad \dot{N_i}= N_i (1 - N_i - \sum_{j \neq i} \alpha_{ij} N_j + h_i(t))
\label{eq:rLV}
\end{equation}
where we have added an external field $h_i$ that will be necessary to define the response of the system to a perturbation. The initial conditions are sampled from a product measure: $P\{N_i(t=0)\}=\prod_{i=1}^SP(N_i(t=0))$. For instance, we generally use a uniform distribution in $[0,1]$ for simulation purposes.

The main steps of the derivation are the following:
\begin{enumerate}
\item For given parameters $\mu$, $\sigma$, $\gamma$ and system size $S$, consider a system whose interactions and initial conditions are drawn for the $N_{i=1..S}$ species;
\item Following the dynamics according to equations \eref{eq:rLV} defines the trajectories $N_{i=1..S}(t)$;
\item We add a new species $N_0$, and therefore draw its initial condition $N_0(0)$ and the interactions $\alpha_{i0}$ and $\alpha_{0i}$ for $i=1..S$;
\item If $S$ is large enough, the impact of this new species on the previous trajectories is a small perturbation and therefore we only consider linear response for the trajectories $\tilde{N}_{i=1..S}(t)$ in the presence of species `0':
$$\tilde{N}_i(t)=N_i(t) - \sum_{j =1..S} \int_0^t \left. \frac{\delta N_i(t)}{\delta h_j(s)}\right|_{h=0} \alpha_{j0} N_0(s) ds$$
The partial derivative are to be understood in a functional sense. We introduce the notation $\chi_{ij}(t,s)=\left.\frac{\delta N_i(t)}{\delta h_j(s)}\right|_{h=0}$
\item We plug these new trajectories in the equation for $N_0$:
$$\dot{N_0}=N_0 (1 - N_0 - \sum_{i \neq 0} \alpha_{0i} \tilde{N}_i + h_0(t))$$
We introduce the matrix $a_{ij}$: $\alpha_{ij}=\mu/S+\sigma a_{ij}$, so that $a_{ij}$ is a Gaussian with zero mean and $1/S$ variance, verifying in addition $\overline{a_{ij}a_{ji}}=\gamma/S$. All the sums $\sum_i$ stand for $\sum_{i=1..S}$, so the interaction term reads:
\begin{eqnarray}
\fl \sum_j \alpha_{0j} \tilde{N}_j=\frac{\mu}{S} \sum_i {N}_i(t)  -\frac{\mu}{S} \sum_{ij} \int_0^t \chi_{ij}(t,s) \left(\frac{\mu}{S} +\sigma a_{j0}\right)  N_0(s) ds \nonumber \\ + \sigma \sum_i a_{0i} {N}_i(t) - \sigma \sum_{ij} a_{0i} \int_0^t \chi_{ij}(t,s) \left( \frac{\mu}{S}+\sigma a_{j0} \right) N_0(s) ds
\label{eq:detailEq}
\end{eqnarray}

\item We take the large S limit and analyze the statistical properties of all terms. The main idea is that by construction $N_{i=1..S}$ are independent from $\alpha_{i0}$ and $\alpha_{0i}$, therefore one can use central-limit-like arguments. Henceforth, the notation $\langle .\rangle$ refers to the average over the couplings $a_{ij}$ and initial conditions $N_i(0)$. We will detail the procedure for the response function term as an example. We start from $\sum_{ij} a_{0i} \chi_{ij}(t,s) a_{j0}$. We consider that the different $\chi_{ij}(t,s)$ are random functions that will depend on the initial conditions $N_{i>0}(0)$ and the interaction matrix $a_{ij>0}$, but are otherwise independent from $a_{j0}$ and $a_{0i}$. 
We first treat the diagonal part. According to the central limit theorem and up to second order contribution, the term $\sum_{i} a_{0i} \chi_{ii} a_{i0}$ will converge towards its average:
$$S \langle  \chi_{ii} a_{i0} a_{0i}\rangle = S \langle  \chi_{ii}\rangle\langle  a_{i0} a_{0i}\rangle=\gamma\langle  \chi_{ii}\rangle$$
We now focus on the non-diagonal part. Its average is zero because $\langle a_{0i} a_{j0}\rangle_{i \neq j}=0$. To determine the scaling of its fluctuations we evaluate the variance of its single components obtaining $\langle \chi_{ij}^2\rangle_{i \neq j} \langle a_{j0}^2 a_{0i}^2\rangle_{i \neq j}$. It can be shown by perturbation theory in the strength of interactions that $\chi_{ij}$ is of order $S^{-1/2}$ for $i\neq j$ \cite{parisi_spin} (see \ref{si:scaling}). Regrouping the scalings, we obtain that $\sum_{i\neq j} a_{0i} \chi_{ij} a_{j0}$ behaves as:
\begin{eqnarray*}
\fl \ S(S-1)\langle \chi_{ij}\rangle_{i \neq j}\langle a_{j0} a_{0i}\rangle_{i \neq j} + \sqrt{S(S-1)}\sqrt{\langle \chi_{ij}^2\rangle_{i \neq j}}\sqrt{\langle a_{j0}^2 a_{0i}^2\rangle_{i \neq j}}Z \\
\sim 0 + S\frac{1}{\sqrt{S}}\frac{1}{S}Z 
\end{eqnarray*}
where $Z$ is a centered standard Gaussian. This shows that the non-diagonal term induces corrections of order $S^{-1/2}$ and can therefore be neglected in the large-S limit. After careful evaluation of all terms in equation \eref{eq:detailEq} according to the same procedure, we get:

\begin{equation*}
\fl \dot{N_0}=N_0 \{ 1 - N_0 - \mu \langle  N_i(t)\rangle  - \sigma \eta(t) \\ + \gamma \sigma^2 \int_0^t \langle  \chi_{ii}(t,s)\rangle N_0(s) ds + h_0(t) \}
\end{equation*}

where $\eta(t)$ is a Gaussian noise with zero mean and covariance $\langle \eta(t)\eta(s)\rangle_\eta = \langle  N_i(t) N_i(s)\rangle$. 

\item Since nothing differentiates $N_0$ from any other species, we obtain the self-consistent equation that leads to dynamical mean field theory:

\begin{equation}
\dot{N}=N \{ 1 - N - \mu m(t)  - \sigma \eta(t) + \gamma \sigma^2 \int_0^t \chi(t,s) N(s) ds + h(t) \}
\end{equation}

where $\eta$ is a Gaussian noise with zero mean and correlator $C(t,s)$, and $m(t)$, $C(t,s)$ and $\chi(t,s)$ are given functions. They are self-consistently determined with the relations:

\begin{equation}
\left\{
\eqalign{
	m(t)&=\mathbb{E}[N(t)] \cr
	C(t,s)&=\mathbb{E}[N(t)N(s)] \cr
	\chi(t,s)&=\mathbb{E}[\left.\frac{\delta N(t)}{\delta h(s)}\right|_{h=0}]
}
\right.
\label{eq:closure}
\end{equation}
In these definitions, the averages $\mathbb{E}[.]$ are now taken with respects to the noise trajectories $\eta$ and the initial condition $N(0)$. Therefore, the equation is self-consistent in the three following functions: the \textbf{average population} $m(t)$, the \textbf{correlator} of the noise $C(t,s)$ and the \textbf{averaged response function} $\chi(t,s)$.

\end{enumerate}

To sum up, we started from an $S$-body deterministic system of differential equations, and ended up with a one-body stochastic self-consistent differential equation \footnote{The derivation is similar to the one of the Langevin equation from Newtonian dynamics \cite{zwanzig2001nonequilibrium}, with the extra-ingredient that the bath corresponds to the rest of the system whose behavior can 
be self-consistently obtained from the one of $N_0$.}. 
It has been mathematically proven \cite{benarous_convInLaw} for spin glasses that when $S\to \infty$, there is a convergence in law between the statistics  of the two descriptions.  We expect that this holds true for our class of models as well, due to the similarity of both the equations and the method.

An important additional remark is that the DMFT is valid as long as we consider times that do not diverge with system size $S$. Otherwise, one cannot neglect terms vanishing with $S$ as we did.

\subsection{DMFT equation for a general class of models}
\label{sec:DMFTGen}

The derivation above can be performed almost identically in more general cases. The only additional subtlety is that we use the fact that the correlation $\langle N_i(t)N_j(t) \rangle$ scales as $S^{-1/2}$ for $i\neq j$, as can be shown by perturbation theory in the strength of interactions \cite{parisi_spin} (see \ref{si:scaling}). Below, we just present the result for a general class of dynamics with a generic and species-dependent response function $R_i(N_i)$, non-linear $p$-body interactions due to $I_i(N_i)$, $J(N_j)$ and a species scaled thermal noise $f_i(N_i)\xi_i(t)$. 

\begin{equation}
\fl \dot{N_i} = R_i(N_i) + I_i(N_i) \left( \sum_{1\leq j_2 < ,.. < j_p \leq S} \alpha^i_{j_2,..j_p} J(N_{j_2})..J(N_{j_p}) + h_i(t) \right) + f_i(N_i)\xi_i(t)
\label{eq:generalModel}
\end{equation}
where $\xi_i$ is a Gaussian white noise, with variance $2\omega^2$. The $i$-dependence of the functions denotes the possible presence of random parameters for each species. For instance, in the general Lotka Volterra case, $R_i(N_i)= r_i/K_i(K_i-N_i)$ where the $r_i$ and $K_i$ respectively correspond to species-dependent growth rates and carrying capacities, that we will treat as random variables sampled from given distributions. The coupling tensor satisfies $ \alpha^i_{j_1,..j_p}=0$ if there exists $k$ such that $i=j_k$, so as not to interfere with the self-interaction $R_i(N_i)$. Otherwise, its cumulants are taken as:

$$\overline{\alpha^i_{j_1,..j_p}}=\mu\frac{p!}{2S^{p-1}} \qquad \overline{(\alpha^i_{j_1,..j_p})^2}_{con}= \sigma^2 \frac{p!}{2S^{p-1}}  \qquad \overline{\alpha^i_{j_1,..j_p} \alpha^{j_k}_{j_1..i..j_{k-1}j_{k+1}..j_p}}_{con}=\gamma \sigma^2 \frac{p!}{2S^{p-1}}$$
where the notation $\overline{X}_{con}$ denotes the connected average of $X$, {\it i.e.} when subtracting their average to the elements. Because of the constraint $1\leq j_1 < ,.. < j_p \leq S$, when considering the cross correlation, there is only one place for the upper index $i$ to go down.

Within this set-up, the DMFT equation for a given species reads:
\begin{eqnarray}
\fl \dot{N}=R_i(N) +I_i(N)\left( \mu m + \sigma \eta + \gamma \sigma^2 \frac{p(p-1)}{2} \int_0^t \chi(t,s) C(t,s)^{p-2} J(N(s)) ds + h \right) \nonumber \\
 + f_i(N)\xi
\label{eq:DMFT_all}
\end{eqnarray}
where $\eta$ is a Gaussian noise with zero mean and covariance $\frac{p}{2} C(t,s)^{p-1}$, and $\xi$ is a Gaussian white noise, with variance $2\omega^2$. Using subscripts for the different times, we obtain the self-consistent closure:

\begin{equation}
\left\{
\eqalign{		
	m(t)&=\mathbb{E}[J(N_t)]^{p-1}\\
	C(t,s)&=\mathbb{E}[J(N_t)J(N_s)]\\
	\chi(t,s)&=\mathbb{E}[J'(N_t)\left.\frac{\delta N_t}{\delta h_s}\right|_{h=0}]
}
\right.
\end{equation}
where the average $\mathbb{E}[.]$ is now taken with respects to the initial condition distribution, the distribution of species-dependent parameters in the functions $R_i$, $I_i$ and $f_i$, the noise trajectory $\eta$ and the thermal noise $\xi$.

It should be stated that the DMFT we derived with the dynamical cavity technique can also be obtained using generating functional technique of Martin-Siggia-Rose-DeDominicis-Janssen \cite{galla_replicator, galla_rLV, eissfeller_numDMFT}.

\subsection{Solving numerically the DMFT equation}
\label{numSolve}

It is difficult to solve numerically a self-consistent equation where the self-consistency applies to functions. We implemented a strategy which works as pictured in \fref{fig:numScheme}. In this section, we write down in details the methodology of the algorithm. The different steps of the program are the following:

\begin{figure}[hbtp]
\centering
\includegraphics[scale=0.5]{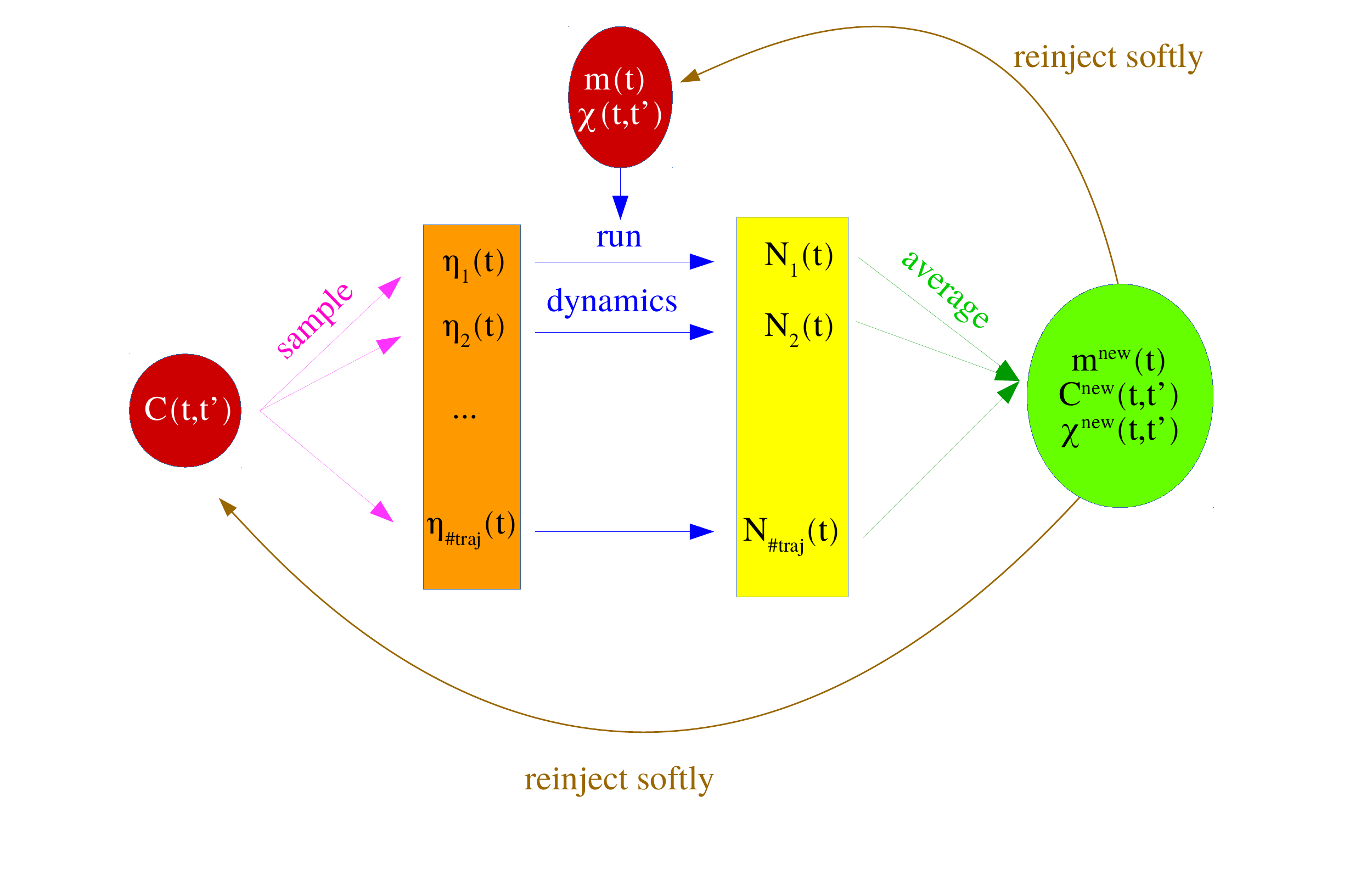}
\caption{Sketch of the numerical scheme for solving the DMFT equation.}
\label{fig:numScheme}
\end{figure}

\begin{enumerate}
\item We start from initial guesses for the correlator matrix $C(t,s)$, the average vector $m(t)$, and the response matrix $\chi(t,s)$. For instance we tried a random average vector for $m$,  diagonal or random positive symmetric matrices for $C$, and lower triangular random matrices for $\chi$ (since $\chi$ is a causal function);
\item Using the correlator, we can sample a Gaussian path as a simple multivariate Gaussian random variable with covariance matrix $p/2C(t,s)^{p-1}$. We draw many ($\#_{traj}$) such Gaussian paths.
\item For each path, we use our guesses $m(t)$ and $\chi(t,s)$ to numerically integrate the DMFT equation where the initial condition is sampled according to the wanted distribution. We used the uniform measure on $[0,1]$ for example. For each Gaussian path, we get a different population trajectory.
\item From these trajectories, we compute the updated values of the average population vector, the correlator matrix and the response matrix (see below), using the self-consistent closure: 

\begin{equation*}
\left\{
\eqalign{
	m^{new}(t)&=\mathbb{E}_{paths}\left[ J(N_t) \right] ^{p-1}\\
	C^{new}(t,t')&=\mathbb{E}_{paths} \left[ J(N_t)J(N_t') \right]\\
	\chi^{new}(t,t')&=\mathbb{E}_{paths} \left[ J(N_t)\int ds \; C^{-1}(t',s)\; \eta(s)\right] \mathbf{or} \; \mathbb{E}_{paths} \left[ \chi_i(t,t')\right]
}
\right.
\end{equation*}

\item We update softly the set of functions: $X^{updated}=(1-a)X+aX^{new}$ with $X$ being respectively $m$, $C$ and $\chi$. The soft reinjection is necessary for the algorithm to converge, and not jump erratically from functions to functions. A reinjection parameter $a=0.3$ seems to be a good choice.

\item We start a new iteration of the loop, with the updated set of functions.

\end{enumerate}

\begin{figure}[hbtp]
\centering
\includegraphics[width=.6\linewidth]{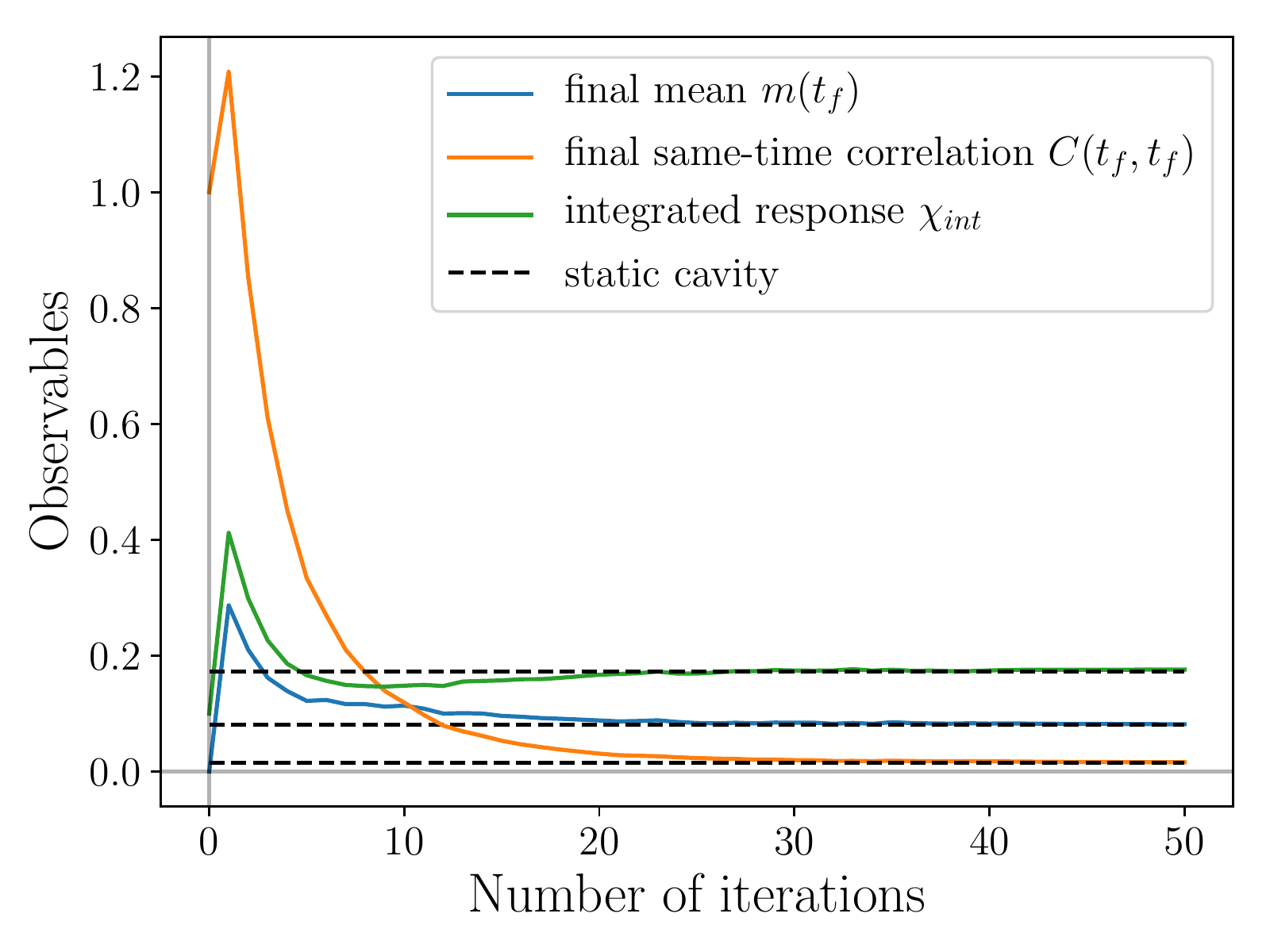}
\caption{We show the convergence of different observables as a function of the number of iterations: the final mean abundance $m(t_f)$, the final same-time correlation $C(t_f,t_f)$ and the integrated response kernel $\chi_{int}=\int_0^{t_f} ds \; \chi(t_f,s)$. For comparison, dotted black lines represent the analytic stationary cavity solutions. The relative errors to the stationary cavity solutions are below $2\%$. As the DMFT observables are computed from a finite number of trajectories $\#_{traj}$, there is always some residual fluctuations. The solver was run with rLV DMFT with parameters $(\mu,\sigma,\gamma,\lambda)=(10,1/2,-1,10^{-4})$ in the Unique Equilibrium phase. The parameters of the program are: reinjection rate $a=0.3$, final time $t_f=40$, discrete time steps $\tau=0.1$, final number of trajectories to average upon $\#_{traj}=10^5$.}
\label{fig:convergenceObs}
\end{figure}

The convergence of the algorithm is of exponential form in the number of iterations, and is independent of the initial set of functions. On \fref{fig:convergenceObs}, we show an example of such a convergence.

Now, let us explain point (iv) in more detail. Obtaining the average population vector, and the correlator matrix from the trajectories is a trivial procedure: one just needs to average. Evaluating the response function $\chi$ is instead more tricky. We studied two different complementary, or alternative, procedures. The first one consists in using Novikov's theorem \cite{novikov} (or Stein's lemma) in the statistical field formulation in order to obtain:
\begin{equation}
\label{eq:Girsanov}
\chi(t,s)=\sigma^{-1}\mathbb{E}[J(N_t)\int dx \; C^{-1}(s,x) \; \eta(x)]
\end{equation}
In this formulation, $C^{-1}$ denotes the matrix inverse of $C$. The detailed derivation is presented in \ref{si:girsanov}. This expression is easy to implement, however it is sometimes too greedy for numerics. For instance, in our problem with multiplicative noise, the number of DMFT trajectories to average upon in order to obtain a satisfactory estimate for the response function is too high. We thus derive another relation, by directly applying $\frac{\delta}{\delta h(t')}$ to \eref{eq:DMFT_all}. In this way, for each trajectory $i$, we can compute the response function $\chi_i(t,t')$ \textit{via} temporal integration, and eventually average over trajectories to obtain $\chi(t,t')$. This procedure is less greedy in terms of needed trajectories, however it is of higher numerical complexity in the number of time steps. The details are given in \ref{si:tempIntegResponse}. In \ref{si:responseSchemes}, we sum up and compare the adequacy of the two methods.

The algorithm we presented here can still be improved in several ways. More specifically, when the response function is needed (when $\gamma \neq 0$), the above algorithm is quite expensive numerically. The complexity of the algorithm might be reduced by proceeding in time slices. Indeed, instead of trying to make the observables converge for the whole time interval $[t_0,t_f]$, it should be numerically more efficient to make them converge on $[t_0,t_1]$, then on $[t_1,t_2]$ using the already converged result of $[t_0,t_1]$, and so on.

The details of the numerical implementation are in \ref{si:numStrategy}, and a public gitHub repository with the corresponding Python programs is available \cite{felixroy_implements_2019}.

\subsection{Numerical check of the results}

\begin{figure}
\centering
\begin{minipage}{.5\textwidth}
  \centering
  \includegraphics[width=\linewidth]{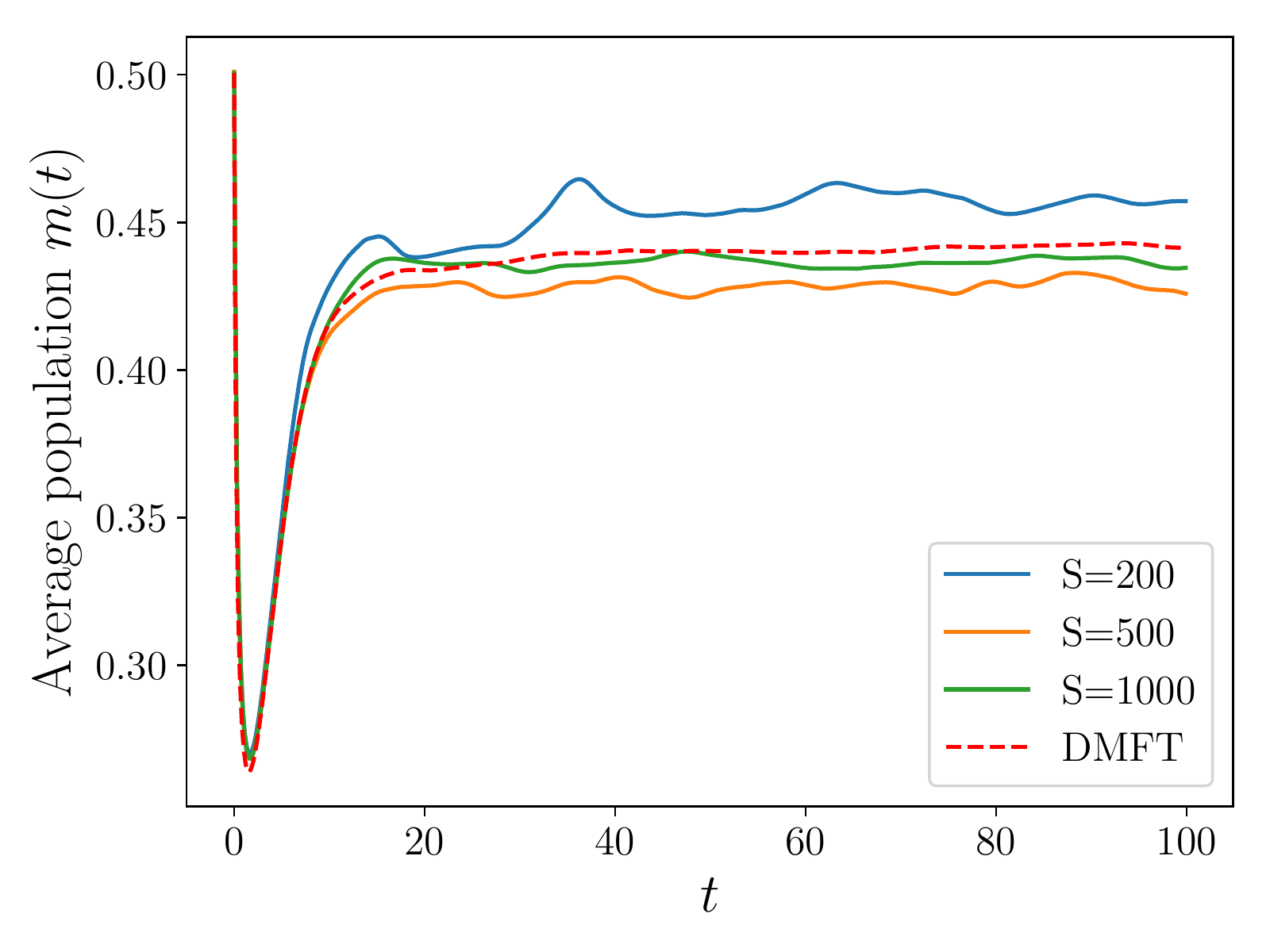}
\end{minipage}%
\begin{minipage}{.5\textwidth}
  \centering
  \includegraphics[width=\linewidth]{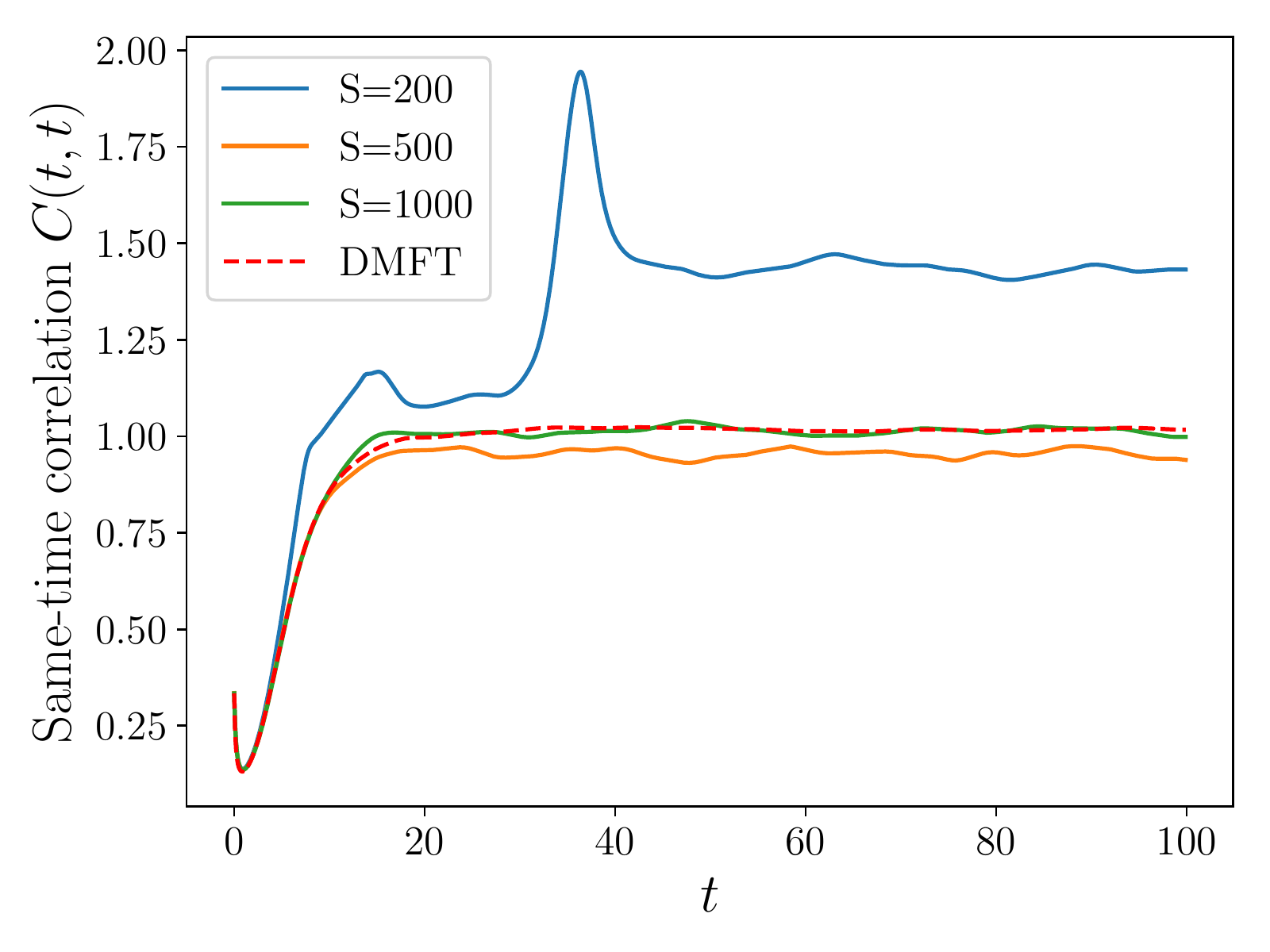}
\end{minipage}
\caption{ Comparison of the observables $m(t)$ (Left) and $C(t,t)$ (Right) between direct simulations varying the ecosystem size $S$, and DMFT predictions in dotted red line. It shows the convergence in law towards DMFT as $S$ increases. The parameters of the simulations are $(\mu,\sigma,\gamma,\lambda) =(4,2,0,10^{-4})$, in the Multiple Attractors phase.}
\label{fig:compDirectSim}
\end{figure}

We checked that the numerical solution of DMFT is consistent with the one from direct simulations. More specifically, we sample $\#_{instances}=200$ interaction matrices and initial conditions for an ecosystem of size $S$, run the deterministic dynamics, and aggregate the observables by averaging over the $S$ species and $\#_{instances}$ realizations. This is what we call direct simulations. In the Unique Equilibrium phase, the agreement is excellent. In \fref{fig:compDirectSim}, we show the comparison between DMFT and direct numerical simulations in the Multiple Attractors phase. As $S$ increases the direct simulations observables converge at all times to the one from DMFT. It is surprising however that the direct simulations are so different from DMFT for $S=200$. We reckon it is related to the fact that at finite $S$, when sampling the interaction matrix with parameters in the Multiple Attractor phase, there is a non-zero probability to get an interaction matrix that describes an Unbounded Growth ecosystem. This problem makes it difficult to have clean data using direct simulations, and underlines the relevance of DMFT analysis.

\section{Application to the random Lotka-Volterra model}

From the general case of equation \eref{eq:DMFT_all}, we recover the random Lotka-Volterra DMFT by taking:
\begin{equation*}
\left\{
\eqalign{R_i(x)&=x(1-x)+\lambda\\
		I_i(x)&=-x\\
		J(x)&=x\\
		f_i(x)&=0
}
\right.
\end{equation*}

From this we obtain: 
\begin{equation}
\dot{N}=N \{ 1 - N - \mu m(t)  - \sigma \eta(t) + \gamma \sigma^2 \int_0^t \chi(t,s) N(s) ds + h(t) \} + \lambda
\label{eq:DMFTrLV}
\end{equation}
The self-consistent closure is that of Eq. \eref{eq:closure}. In this section, we show how to get back the stationary results \cite{buninPRE} from DMFT, and we study the stability of such stationary solution. All the following results are valid for $\lambda>0 $ where the limit $\lambda \to 0$ is subsequently taken. 

\subsection{How to recover the stationary results}
If the ecosystem parameters belong to the Unique Equilibrium phase, each species will eventually reach a final population value and stops changing. We describe this final state using DMFT. The one-species stochastic process becomes time-independent, so the derivative is zero, the average $m(t)$ converges to a number $m(\infty)$, the population $N(t)$ and the Gaussian noise $\eta(t)$ converge to random variables $N(\infty)$ and $\eta(\infty)$. As the process is stationary, we treat the memory kernel as time-translational invariant $\chi(t,s)=\chi(t-s)$ and therefore:
$$\int_0^t \chi(t,s) N(s) ds=\int_0^t \chi(u) N(t-u) du \to_{t\to \infty} \int_0^\infty \chi(u) du N(\infty)$$
Introducing the integrated memory kernel $\chi_{int}=\int_0^\infty du \; \chi(u)$, the DMFT equation \eref{eq:DMFTrLV} finally converges to:
\begin{equation}
0=N_\infty \{ 1 - N_\infty - \mu m_\infty  - \sigma \eta_\infty + \gamma \sigma^2 \chi_{int} N_\infty + h \}
\label{eq:DMFT1eq}
\end{equation}
\begin{equation}
\label{eq:closEqRS}
\left\{
\eqalign{
		m_\infty&=\mathbb{E}[N_\infty]\\
		\chi_{int}&=\mathbb{E}[\frac{\delta N_\infty}{\delta h}]\\
		\mathbb{E}[\eta_\infty^2]&=\mathbb{E}[N_\infty^2]
}
\right.
\end{equation}

From equation \eref{eq:DMFT1eq}, $N_\infty$ can either be 0 or $(1-\gamma \sigma^2 \chi_{int})^{-1}(1 - \mu m_\infty  - \sigma \eta_\infty )$. By a simple linear stability analysis performed on the real system of $S$ species (see \ref{siSec:stability_dead}), it can be shown that the 0 solution is linearly unstable when the other solution is positive. Therefore, we obtain:

\begin{equation}
N_\infty = \max \left( 0,  \frac{1 - \mu m_\infty  - \sigma \eta_\infty }{1-\gamma \sigma^2 \chi_{int}} \right)
\label{eq:rsGuy}
\end{equation}
so the random variable $N_\infty$ follows a Gaussian distribution, truncated for negative abundances. We write the closed system of equations in \ref{siSec:1eq}, and show that we end up with the same system as the one from \cite{buninPRE}. From it, all observables can be computed numerically as a function of the parameters $(\mu,\sigma,\gamma)$: the fraction of species coexisting at the fixed-point, the mean abundance $N$ of species that survive and the mean response function. Some further analytical results can be derived as well, such as identifying in parameter space the boundary between the Unique Equilibrium phase and the Unbounded Growth phase. However, this analysis is only exact  when we are in the Unique Equilibrium phase. It becomes approximate in the Multiple Attractors phase.

\subsection{Dynamical stability and the transition line to Multiple Attractors}

We now describe the loss of stability of the Unique Equilibrium solution when increasing the variability $\sigma$ of interactions: a dynamical phase transition takes place. The setup follows the one of \cite{opperPRL}: starting in the Unique Equilibrium phase, we let the system reach an equilibrium point, then add some small field $h(t)$ which we will take as a Gaussian white noise with covariance $\overline{h(t)h(s)}=\sigma_h^2\delta(t-s)$, and see how the system responds in perturbation theory. In order to do so, we consider the DMFT equation \eref{eq:DMFTrLV}, linearize it around a stationary solution, and perform a Fourier analysis \cite{opperPRL}. The detailed calculations are presented in \ref{si:stability_1eq}. Introducing $\tilde{X}$ the Fourier transform of $X$, we obtain the small frequency expansions for both the connected correlator $C_c(t,t')=\mathbb{E}[\eta(t)\eta(t')] - \mathbb{E}[\eta_\infty^2]$ and the response function:

\begin{eqnarray}
\label{eq:smallOmExpC}
&\tilde{C_c}(\omega) = \frac{\sigma^2_h}{ \frac{(1-\gamma\sigma^2 \chi_{int})^2}{\phi} -\sigma^2 +|\omega| \frac{|\chi_{int}|}{\phi^2} \pi p_+(0)}\\
&\tilde{\chi}(\omega) = - \frac{\phi}{1-\gamma \sigma^2 \chi_{int}} +i|\omega| \log|\omega| \frac{p_+(0)}{(1-\gamma \sigma^2 \chi_{int})^2 - \gamma \sigma^2 \phi} 
\label{eq:smallOmExpK}
\end{eqnarray}
where $\phi$, $\chi_{int}$ and $p_+(0)$ are properties of the Unique Equilibrium we started from. They correspond respectively to the fraction of surviving species, the integrated response to perturbations, and the value in $0^+$ of the surviving species' distribution. They can be computed using the stationary cavity equations in \ref{siSec:1eq}.

Different things should be noted about these expansions. First, it can be checked that we obtain the same zeroth order condition for the response function as in the stationary cavity study: $\tilde{\chi}(\omega=0)=\chi_{int}$. Secondly, the correlator initially behaves as $(a+|\omega|)^{-1}$ which corresponds to a temporal decay as $1/t^2$. But a change of behaviour is displayed when zeroth order term $a$ goes to zero: we observe a $1/f(\sim1/\omega)$ correlation spectrum, which is an indicator of the chaotic transition \cite{opperPRL}. Indeed, with this criterion we find the same transition in parameter space as the one from random matrix theory (the line $\sigma_c=\frac{\sqrt{2}}{1+\gamma}$ in the phase portrait in \fref{fig:phaseP}). Surprisingly, the response function instead does not exhibit a transition at $\sigma_c$, except for $\gamma=1$ where the fluctuation-dissipation theorem establishes a direct link between the correlation function and the response function. More complex response functions might be needed to locate the transition in the general case.

\section{Numerical solution for the random Lotka-Volterra model}

In this section, we present some numerical results for the random Lotka-Volterra DMFT, and show the consistency of both analytics and numerics. The aim is to illustrate the quality of the DMFT results, and present a first description of the dynamical phases (a more complete one will be presented elsewhere).  

\subsection{Result in the Unique Equilibrium phase}

\begin{figure}[hbtp]
\centering
\includegraphics[scale=0.7]{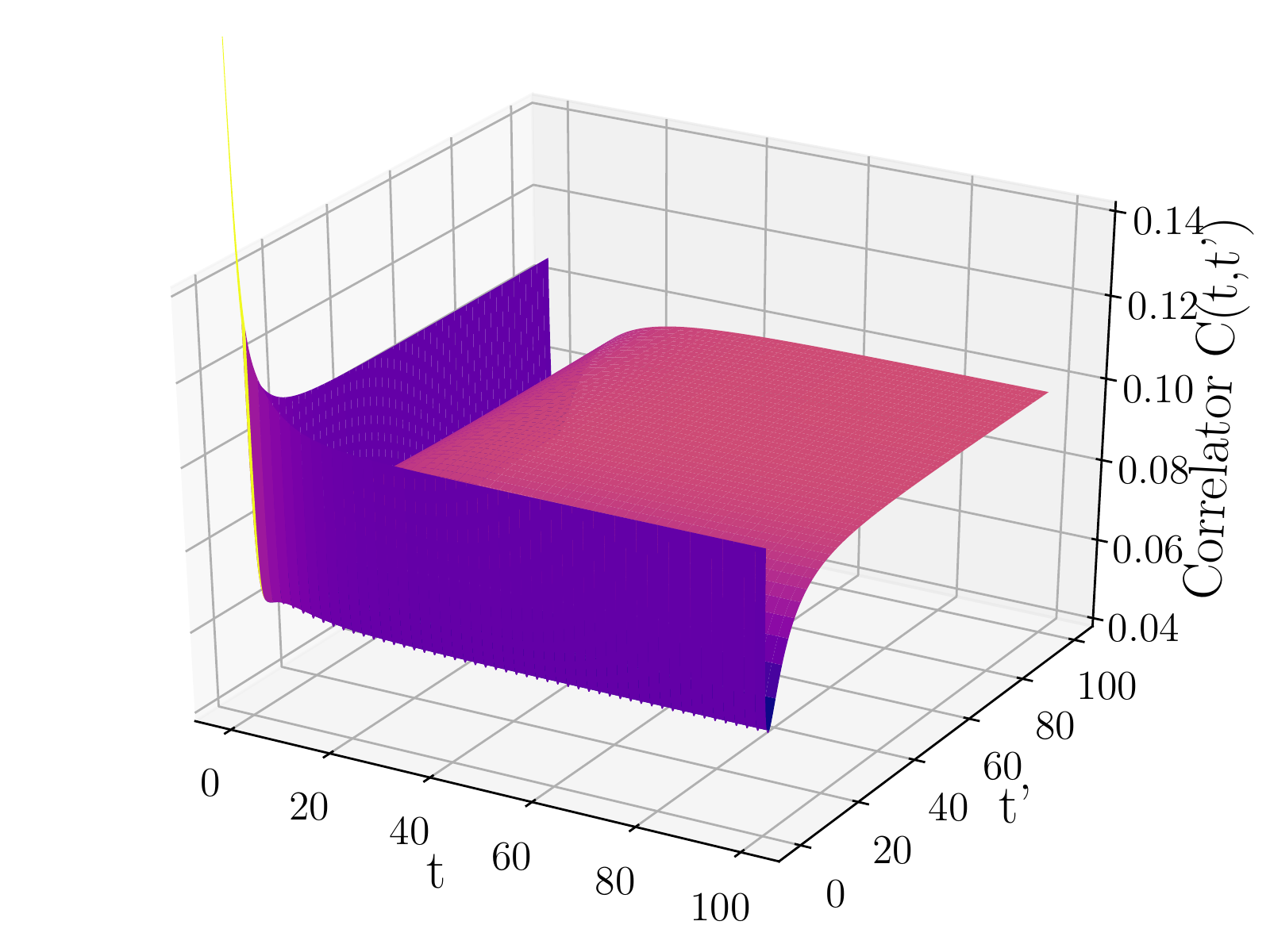}
\caption{Numerical correlator of plateau type, for rLV DMFT with parameters $(\mu,\sigma,\gamma,\lambda)=(4,1,0,10^{-4})$ below the onset of chaos. The parameters of the program are the same as in \fref{fig:convergenceObs}.}
\label{fig:numPlateau}
\end{figure}

We focus on the correlator $C(t,t')=\mathbb{E}[N(t)N(t')]$. In the Unique Equilibrium phase, it reaches a plateau as each trajectory converges to a random constant. Moreover, the value of the plateau coincides with the stationary cavity observable $q$. This is indeed the case, as pictured on \fref{fig:numPlateau}. The convergence to the stationary solution is a good check of the validity of our numerical strategy. It is shown more precisely on \fref{fig:convergenceObs}.

\subsection{Result in the Multiple Attractors phase}

\begin{figure}
\centering
\begin{minipage}{.5\textwidth}
  \centering
  \includegraphics[width=\linewidth]{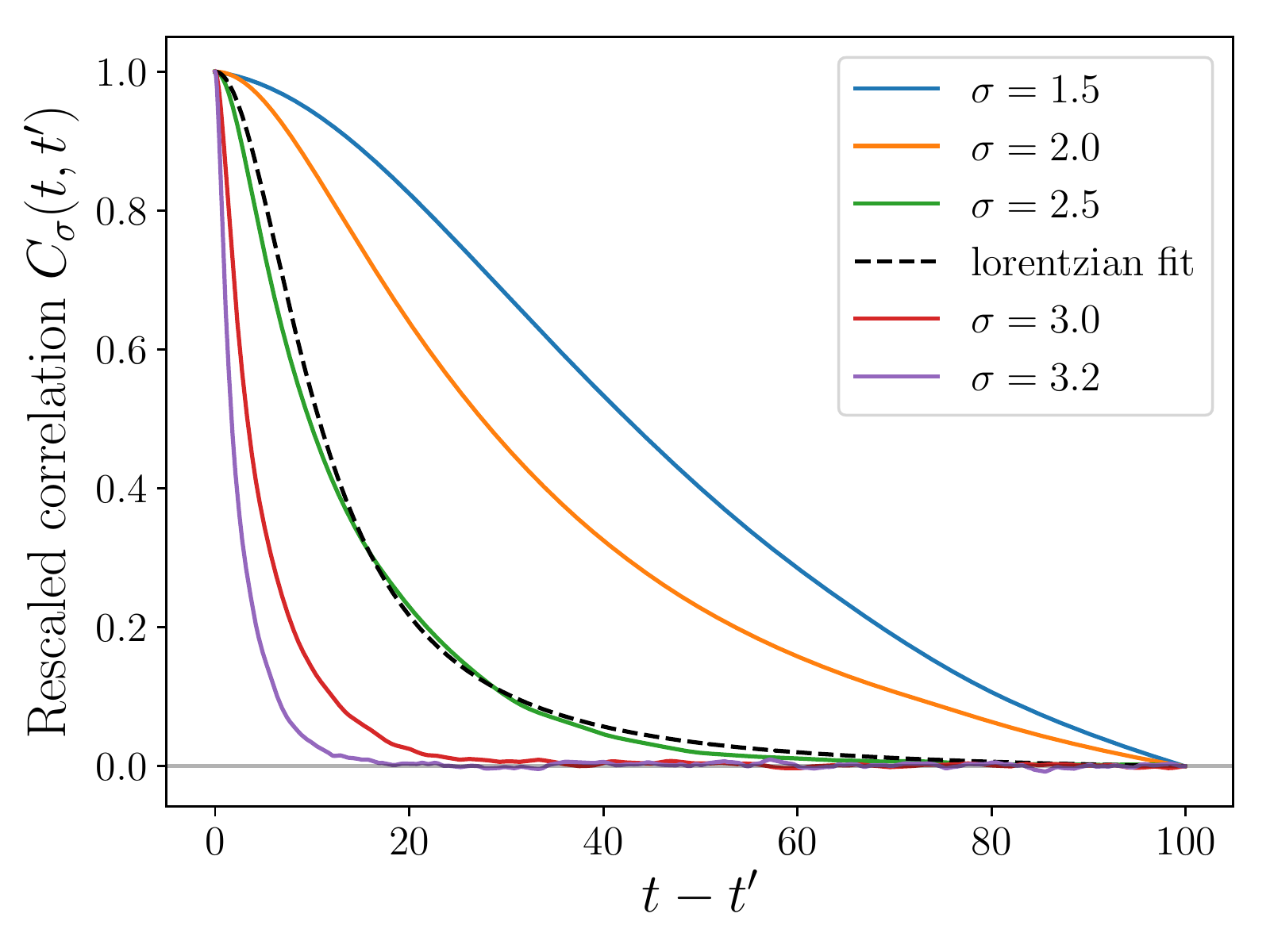}
\end{minipage}%
\begin{minipage}{.5\textwidth}
  \centering
  \includegraphics[width=\linewidth]{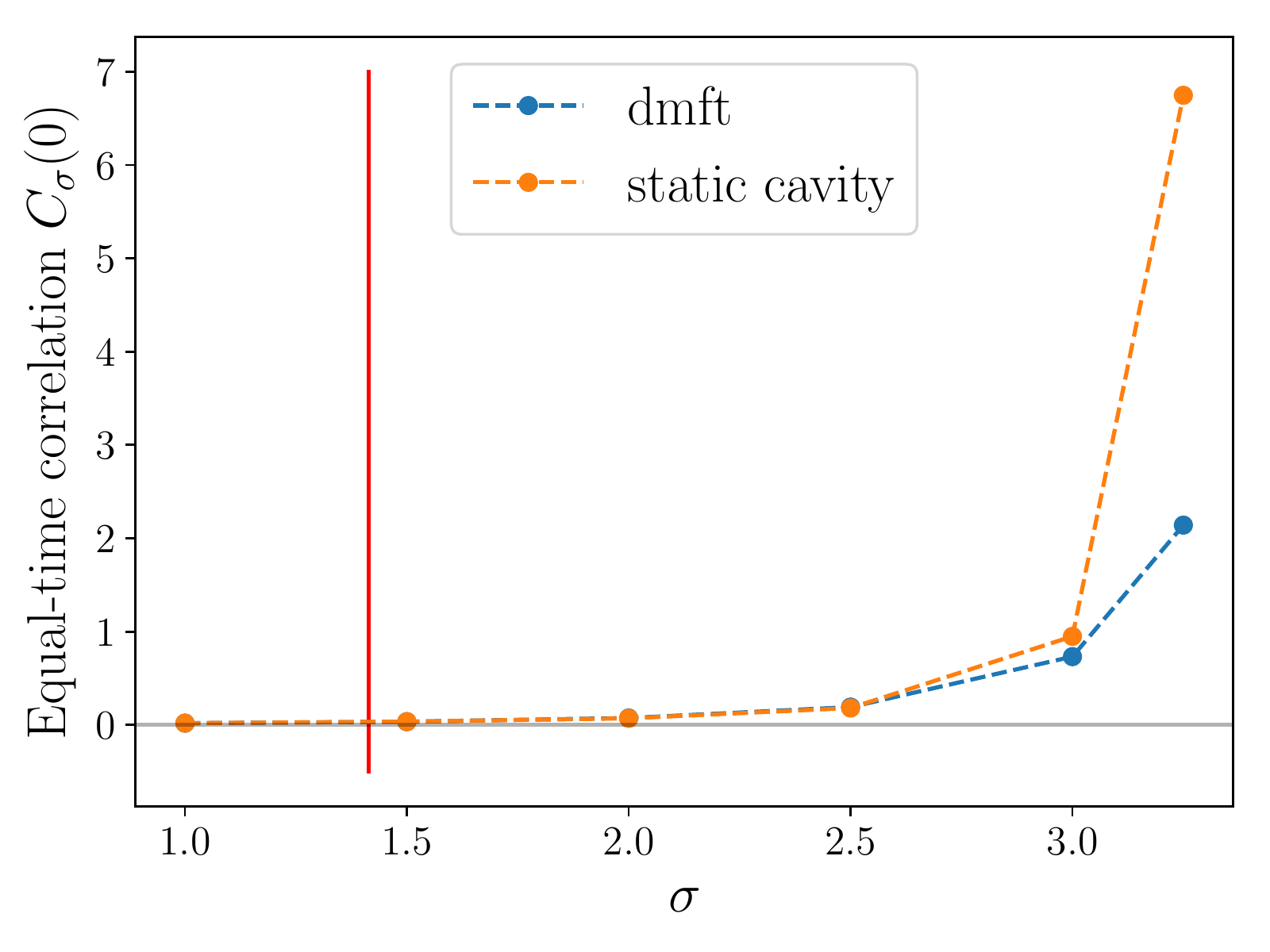}
\end{minipage}
\caption{Left: Time evolution of the TTI correlation $C_\sigma(t-t')$ varying $\sigma$. Those are DMFT numerical results with parameters $(\mu, \gamma, \lambda)=(10,0,10^{-4})$. We checked that the system indeed reaches TTI, $t'=200$ is enough here. More precisely, we show the rescaled TTI correlator $\frac{C_\sigma(t-t')-C_\sigma(\infty)}{C_\sigma(0)-C_\sigma(\infty)}$, in order to see the dependence of the chaotic time scale $a_\sigma$ with $\sigma$. This time scale decreases with $\sigma$. In order to have a quantitative approximation for $a_\sigma$, we use a Lorentzian fit; an example of such is the dotted black curve. Right: $\sigma$ dependence of the TTI equal-time correlator $C_\sigma(0)$. The red line indicates the chaotic transition. In orange dots, we show for comparison the analytical static cavity results. In the Unique Equilibrium phase, the DMFT and static cavity results coincide. In the Multiple Attractors phase, they diverge from each other, but the static cavity remains a good approximation for a relevant chaos depth. Note that $C_{\sigma}(0)>0$ for all $\sigma$.}
\label{fig:chaos_TTI}
\end{figure}

\begin{figure}
\centering
\begin{minipage}{.5\textwidth}
  \centering
  \includegraphics[width=\linewidth]{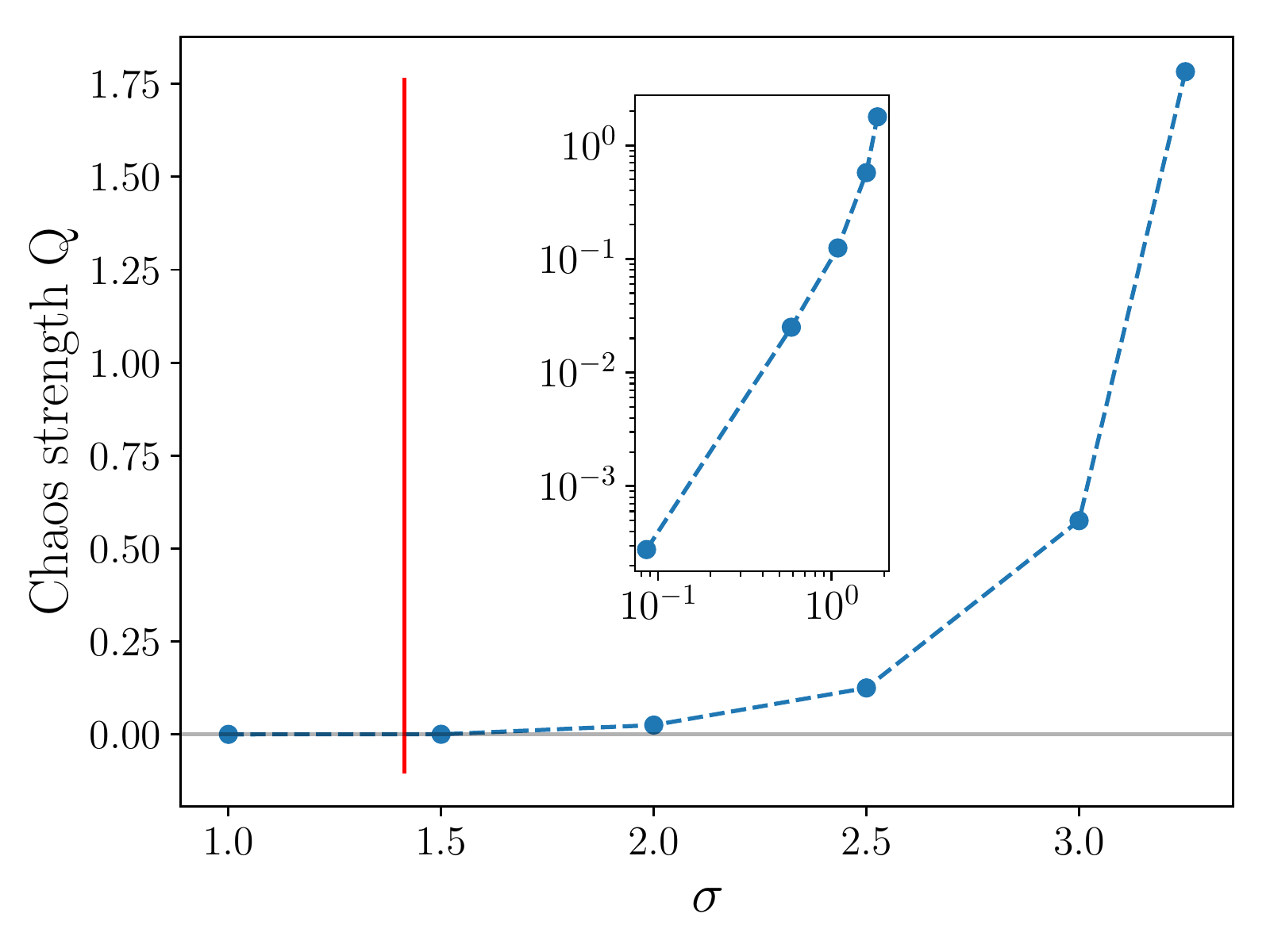}
\end{minipage}%
\begin{minipage}{.5\textwidth}
  \centering
  \includegraphics[width=\linewidth]{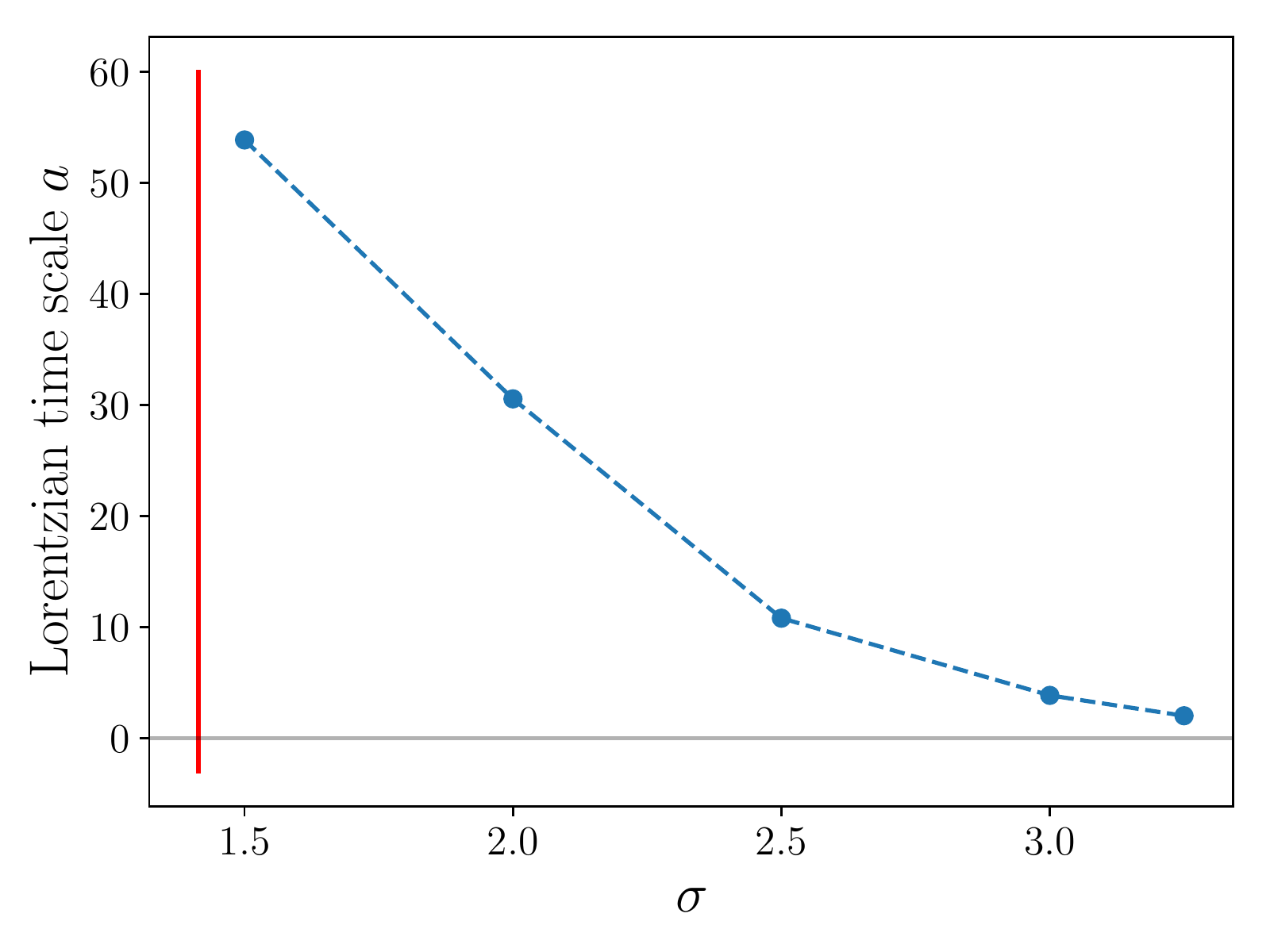}
\end{minipage}
\caption{Left: Chaos strength $Q_\sigma=C_\sigma(0)-C_\sigma(\infty)$ as a function of $\sigma$. It is zero in the Unique Equilibrium phase, and non-zero in the Multiple Attractors phase. The red line corresponds to the chaotic transition $\sigma=\sigma_c$. The inset is a log-log plot $Q(\sigma - \sigma_c)$. The behaviour seems to agree with a critical exponent around 2.4: $Q \sim (\sigma - \sigma_c)^{2.4}$. Right: Chaos time scale $a_\sigma$ as a function of $\sigma$. It is non-zero in the Multiple Attractors phase, and should diverge as we approach the chaotic transition. The red line corresponds to the chaotic transition $\sigma=\sigma_c$. These values are only approximate, based on basic lorentzian fit from \fref{fig:chaos_TTI}. They do not allow us to extract a critical exponent.}
\label{fig:chaos_orderParam}
\end{figure}

In the Multiple Attractor phase we expect a different behaviour. The system does reach a time-translational invariant (TTI) chaotic state. This means that the one-time observables (the mean population $m(t)$, the proportion of alive species $\phi(t)$, or the equal-time correlation $C_{\sigma}(t,t)$) converge to a constant, and the two-time observables become functions of the time difference: $C_\sigma (t,t')=C_\sigma (t-t')$. If we focus on large times, we expect a relaxing behaviour for the correlator, as the trajectory decorrelates from itself when it explores the phase space along the chaotic attractor. We observe this phenomenon in the numerical solutions. Moreover, the TTI state depends on how deep in the Multiple Attractors phase the system is. On \fref{fig:chaos_TTI}, we show the dependence on $\sigma$ of the TTI correlation $C_\sigma (t-t')$, rescaled as follows. These functions $C_\sigma (\tau)$ starts at a TTI value for the equal-time correlation $C_\sigma (0)$, then as the trajectories decorrelate from themselves $C_\sigma(\tau)$ relaxes towards a TTI final value $C_\sigma(\infty)$ over a timescale $a_\sigma$. We therefore plot $\frac{C_\sigma(t-t’)-C_\sigma(\infty)}{C_\sigma(0)-C_\sigma(\infty)}$. We also denote $Q_\sigma=C_\sigma(0)-C_\sigma(\infty)$ the amplitude of the decorrelation. It is representative of the chaos strength, and this is an order parameter for the chaotic transition. On \fref{fig:chaos_orderParam} we show the dependence of both the chaos strength $Q_\sigma$ and the time scale $a_\sigma$ as a function of the chaotic depth $\sigma-\sigma_c$. As expected, the chaos strength $Q_\sigma$ increases and the chaos time scale $a_\sigma$ decreases with the chaotic depth. Our results show that chaos emerges through a {\it second-order out of equilibrium dynamical phase transition}. A first attempt to obtain critical exponents is shown in \fref{fig:chaos_orderParam}. A more thorough study will be presented elsewhere.

We recall that we have considered small but finite immigration. Dynamics without immigration is different, as we discuss below.

\subsection{Aging dynamics without immigration}
\begin{figure}
\centering
\begin{minipage}{.5\textwidth}
  \centering
\includegraphics[width=\linewidth]{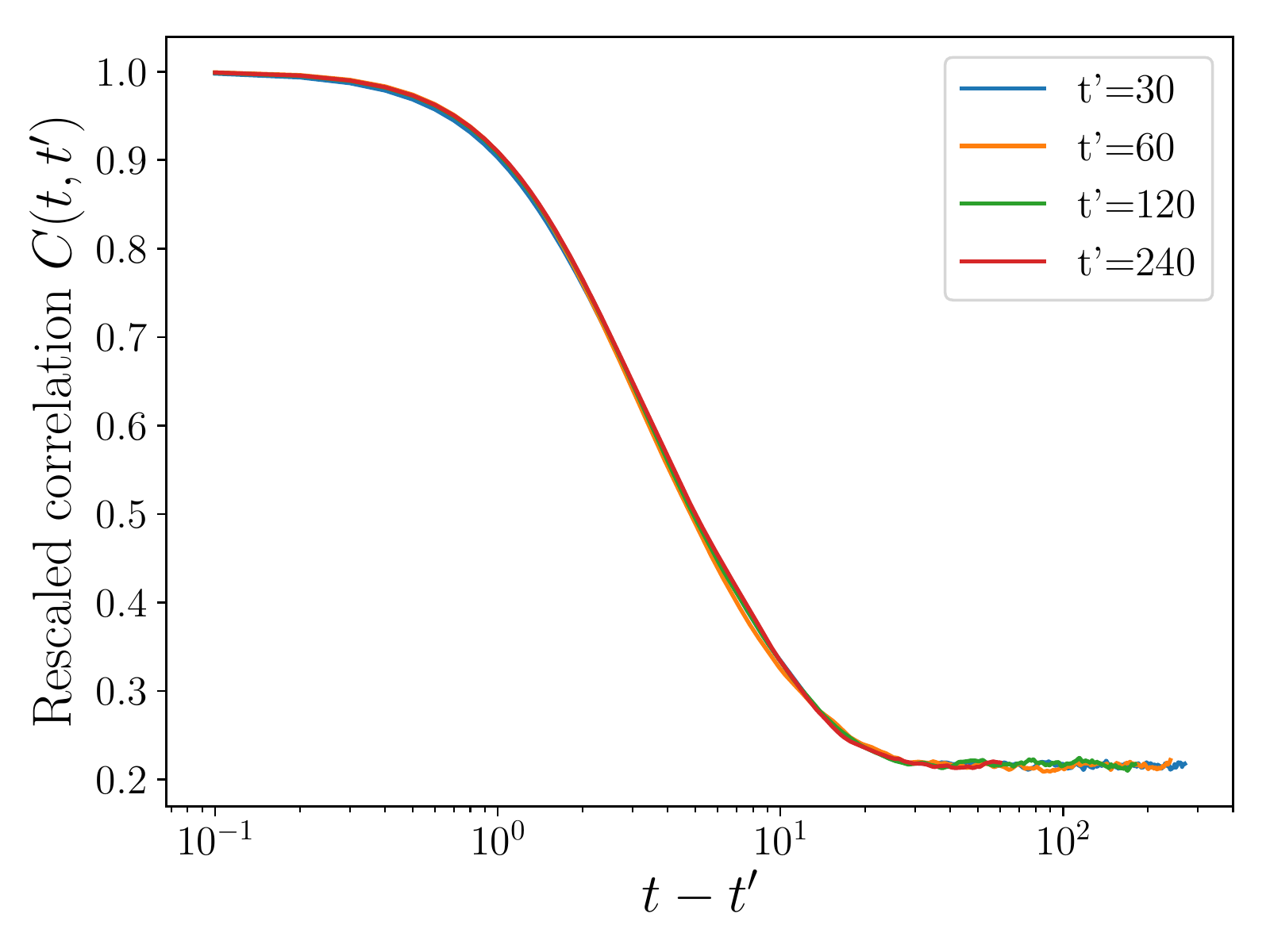}
\end{minipage}%
\begin{minipage}{.5\textwidth}
  \centering
     \includegraphics[width=\linewidth]{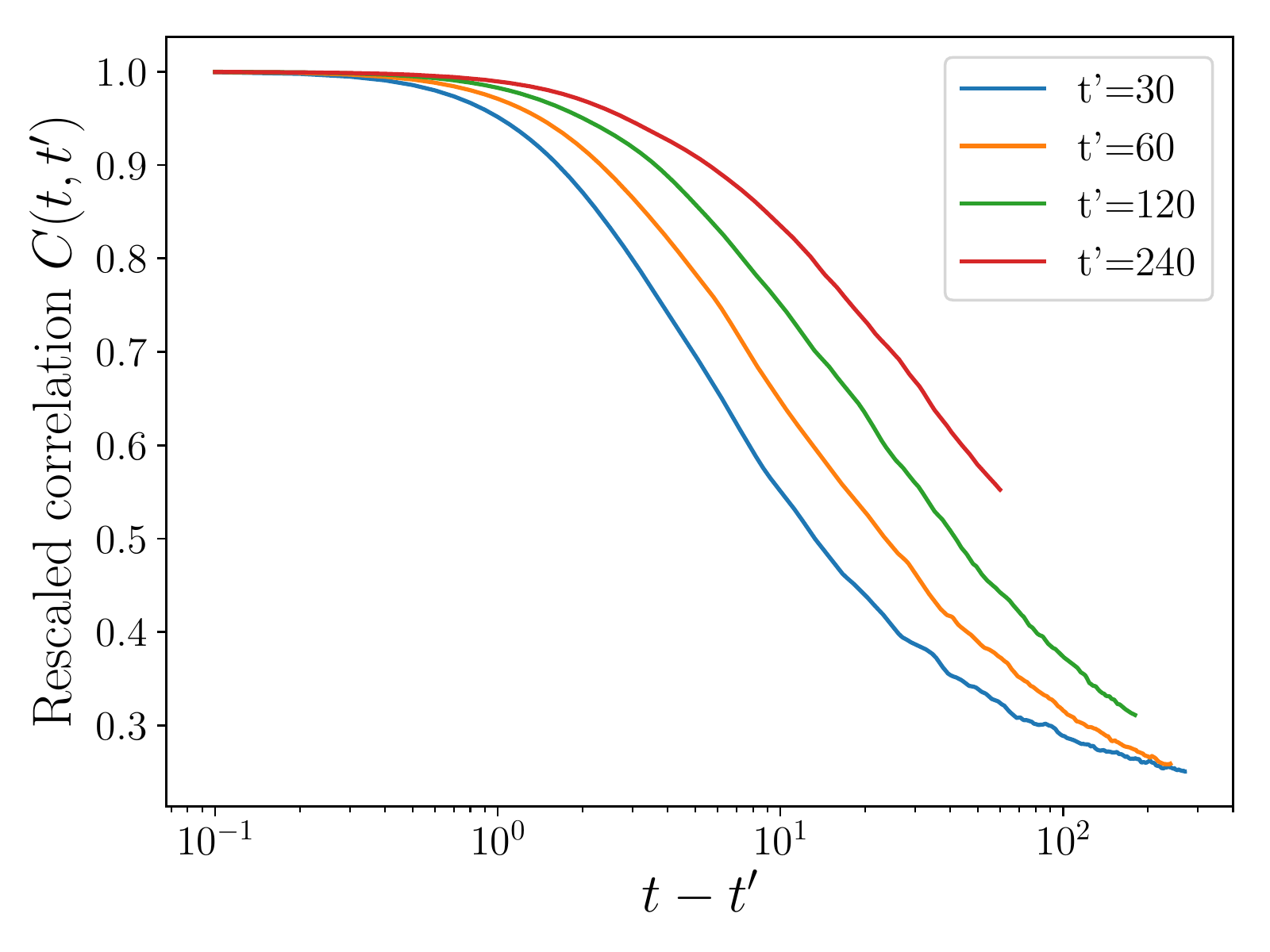}
\end{minipage}
\caption{Aging phenomenon without immigration in DMFT. In the chaotic phase, we show the decay of the rescaled correlation $C(t,t')/C(t',t')$ as a function of $(t-t')$, varying $t'$. The parameters are $(\mu,\sigma,\gamma)=(10,3,0)$ Left: With immigration $\lambda=10^{-4}$, the system reaches a TTI state, there is no dependence on the age of the system $t'$. Right: Without immigration $\lambda=0$, the relaxation of the correlation does depend on the age of the system $t'$; the older the system, the longer it takes to relax. The parameters of the program are the same as in \fref{fig:convergenceObs}.}
\label{fig:aging}
\end{figure}

We now consider the effect of the absence of immigration on the chaotic dynamics. The main issue is that 
chaos induces fluctuations that can drive species to extinction in absence of immigration and, hence, potentially kill chaos itself. The sustainability of chaotic dynamics without immigration is therefore far from being granted, actually a very different dynamical behavior can be present when $\lambda=0$. 
Here we show that this is indeed the case for $\gamma=0$. In \fref{fig:aging} 
we compare the correlation functions, normalized by its equal time value, obtained by DMFT for $\gamma=0$ with and without immigration. In the former case (left panel), it is clear that a stationary chaotic state establishes as $C(t,t')$ becomes a function of $(t-t')$ at large times. On the contrary, without immigration (right panel), $C(t,t')$ shows the aging behavior characteristic of glassy system: the correlation function is not a function of $t-t'$ and displays a relaxation that is slower the older is the system. This is a nice illustration of how our numerical implementation of DMFT allows to unveil the existence of different and complex dynamical behaviors. \\
A detailed understanding of the aging chaotic behavior shown in \fref{fig:aging}, its dependence on the degree of asymmetry $\gamma$, and a thorough analysis of how and when chaos fades away is left for a future work.

\section{Conclusion}
In summary, we have presented a general derivation of DMFT for models of ecosystems based on the dynamical cavity method. 
We have implemented and tested our numerical method for generalized Lotka-Volterra models of ecosystems 
and showed that it can capture complex dynamics such as chaos and aging.
Future works will be devoted to a thorough analysis of these complex dynamical regimes, and also 
to improvements of our algorithm along the ways discussed in this paper.  \\
The main contribution of our work is the development of a numerical method 
to solve DMFT that can be used for many different systems characterized by stochastic dynamics 
and by a large number of degrees of freedom. One important potential application is to the dynamics
of interacting particle glassy systems in the limit of infinite dimensions for which mean-field dynamical 
equations were derived recently \cite{kurchan2016,infDim_agoritsas}. 
\ack

This work was partially funded by Capital Fund Management - Fondation pour la Recherche, and Simons Foundation collaboration
Cracking the Glass Problem (No. 454935 to G. Biroli). G. Bunin acknowledges support by the Israel Science Foundation (ISF) Grant no. 773/18. 
This research was supported in part by the National Science Foundation under Grant No. PHY-1748958.

We thank A. Altieri, J.-P. Bouchaud, D.S. Fisher, T. Galla, A. Manacorda, V. Ros and F. Zamponi for useful interactions on this topic.

\clearpage

\appendix

\section{Scaling of the cross response function and cross correlation}
\label{si:scaling}

\subsection{Cross reponse function}

We start from the simplified rLV case:
$$\forall i =1..S, \quad \quad \dot{N_i}= N_i (1 - N_i - \sum_{j \neq i} \alpha_{ij} N_j + h_i(t))$$

We derive this equation with respect to $h_l(t')$, in the functional sense. We'll denote $\chi_{il}(t,t')=\frac{\delta N_i(t)}{\delta N_l(t')}$. We obtain:

$$\frac{\partial}{\partial t} \chi_{il}(t,t') = \chi_{il}(t,t') \left[ \frac{d}{dt} \log N_i(t) - N_i(t) \right] + N_i(t) \left[ \delta(t-t')\delta_{il} - \sum_j \alpha_{ij} \chi_{jl}(t,t') \right]$$

We know the diagonal response to be of order one at short time, and decaying: $\chi_{ii} \sim 1$. We expect that the response conveyed through correlation loop will be subleading compared to the diagonal response: $\chi_{ii} \gg \chi_{i\neq l}$. Therefore, its scaling can be inferred by considering only the contribution from the diagonal in its time evolution:

$$\frac{\partial}{\partial t} \chi_{i \neq l}(t,t') \sim - N_i(t) \;  \alpha_{il} \; \chi_{ll}(t,t')\rightarrow\chi_{i\neq l} \sim \alpha_{il} \sim S^{-1/2}.$$

\subsection{Cross correlation}

If we had the Fluctuation-Dissipation theorem, we would directly get the scaling of the cross correlation $C_{i\neq l} \sim \chi_{i\neq l} \sim S^{-1/2}$. However, this is not the case, but we show that the scaling relation still holds here. We start again from the simplified rLV case:
$$\dot{N_i}=f(N_i)- N_i \sum_j \alpha_{ij} N_j$$
where $f(N_i)=N_i (1 - N_i)$, and remember that $\alpha_{ii}=0$. We use perturbation theory in the interaction matrix $\alpha = ||\alpha||$. Denoting the solution of the equation $N_i^0$ when $\alpha=0$ and introducing $N_i= N_i^0+\alpha \delta N_i + \alpha^2...$, we obtain the first order correction for two different species $i\neq l$ through linear response:

$$\delta N_i(t) = -\int dt_1 \chi_{ii}(t,t_1) \sum_j \alpha_{ij} N_j^0(t_1)$$
$$\delta N_l(t') = -\int dt_2 \chi_{ll}(t',t_2) \sum_{j'} \alpha_{lj'} N_{j'}^0(t_2)$$

From this relation, we compute the connected averages: 

\begin{eqnarray*}
\fl \langle N_i(t) N_l(t') \rangle_{con} &= \langle \delta N_i(t)\delta N_l(t') \rangle_{con} \nonumber\\
  &= \int dt_1 dt_2 \chi_{ii}(t,t_1) \chi_{ll}(t',t_2) \sum_{jj'} \alpha_{ij} \alpha_{lj'} \langle N_j^0(t_1) N_{j'}^0(t_2) \rangle_{con} \nonumber 
\end{eqnarray*}

We remark that the last term $\langle N_j^0(t_1) N_{j'}^0(t_2) \rangle_{con}$ corresponds to the connected correlation $C_{jj',con}(t_1,t_2)$. Then, from the same argument as the cross response function, we expect the cross correlation to be subleading compared to the diagonal one: $C_{ii,con} \gg C_{i\neq l,con}$. Therefore we only consider the diagonal contributions:

\begin{eqnarray*}
\fl \langle \delta N_i(t)\delta N_l(t') \rangle_{con} &= \int dt_1 dt_2 \chi_{ii}(t,t_1) \chi_{ll}(t',t_2) \sum_{j} \alpha_{ij} \alpha_{lj} C_{jj,con}(t_1,t_2) \nonumber\\
  &\sim \sum_{j} \alpha_{ij} \alpha_{lj} \nonumber 
\end{eqnarray*}

The last term is a random variable with average $\mu^2/S$, and variance $\sigma^4/S$, therefore we obtain $C_{i \neq l} \sim S^{-1/2}$.

%
%
%

\section{Novikov's theorem and generating functional formalism}
\label{si:girsanov}
We want to evaluate: $\chi(t,t')=\mathbb{E}[J'(N_t)\left.\frac{\delta N_t}{\delta h_{t'}}\right|_{h=0}]$. For simplicity, we forget about averaging over initial conditions and thermal noises, it does not change the proof. We introduce the distribution of the population trajectories:
$$\mathbb{P}\{N\}=\int D\eta \; \mathbb{P}\{\eta\} \; \mathbb{P}\{N|\eta,h\}$$
where the brackets denotes functional distributions. $\eta$ is a Gaussian noise, so its probability measure is given up to a normalization factor by:
$$\mathbb{P}\{\eta\} \quad \alpha \quad \exp \left( - \frac{1}{2}\int dt \; ds \; \eta(t) \; C^{-1}(t,s)\; \eta(s) \right)$$
Furthermore, the $\mathbb{P}\{N|\eta\}$ distribution is deterministic and follows the DMFT dynamics. It is therefore a Dirac-distribution:

$$\mathbb{P}\{N|\eta,h\}=\prod_t \delta \left(\dot{N}_t-R(N_t) -I(N_t)\left( .. \sigma \eta + .. + h \right)  + f(N)\xi \right)$$

We now have everything to write down the average:
\begin{eqnarray*}
\chi(t,t')&=\mathbb{E}[J'(N_t)\left.\frac{\delta N_t}{\delta h_{t'}}\right|_{h=0}]\\
&=\frac{\delta}{\delta h_{t'}}\mathbb{E}[J(N_t)]\\
&=\frac{\delta}{\delta h_{t'}} \int DN \; D\eta \; \mathbb{P}\{\eta\} \; \mathbb{P}\{N|\eta,h\} \; J(N_t)\\
&= \int DN \; D\eta \; J(N_t) \; \mathbb{P}\{\eta\} \; \frac{\delta}{\delta h_{t'}} \; \mathbb{P}\{N|\eta,h\}
\end{eqnarray*}
When we write the average as an integration over the different paths, they become non-correlated variables. The correlation aspect is taken care of in the distributions. In addition, the distribution $\mathbb{P}\{N|\eta,h\}$ is symmetric in $h(t')$ and $\sigma \eta(t')$. We also perform an integration by part and find:
\begin{eqnarray*}
\chi(t,t') &= \int DN \; D\eta \; J(N_t)\; \mathbb{P}\{\eta\} \; \frac{\delta}{\sigma \delta \eta_{t'}} \; \mathbb{P}\{N|\eta,h\}\\
&= - \int DN \; D\eta \; J(N_t)\; \mathbb{P}\{N|\eta,h\} \; \frac{\delta}{\sigma \delta \eta_{t'}} \; \mathbb{P}\{\eta\}\\
&= \frac{1}{\sigma} \int DN \; D\eta\; J(N_t) \; \mathbb{P}\{N|\eta,h\} \; \left( \int ds \; C^{-1}(t',s) \; \eta(s) \right) \mathbb{P}\{\eta\}\\
&= \frac{1}{\sigma} \mathbb{E}[J(N_t)\left( \int ds \; C^{-1}(t',s) \; \eta(s) \right)]
\end{eqnarray*}

\section{Temporal integration of the response function}
\label{si:tempIntegResponse}

We remind the DMFT equation \eref{eq:DMFT_all} for a general class of models here, and we will consider each trajectory (denoted with $i$) simulated through this equation:
\begin{eqnarray}
\fl \dot{N_i}=R_i(N_i) + I_i(N_i)\left( \mu m + \sigma \eta_i + \gamma \sigma^2 \frac{p(p-1)}{2} \int_0^t \chi(t,s) C(t,s)^{p-2} J(N_i(s)) ds + h_i \right) \nonumber \\
 + f_i(N_i)\xi_i
\end{eqnarray}

\noindent
We now apply $\frac{\delta}{\delta h_i(t')}$. In this way, for each trajectory $i$, we can compute the response function $\chi_i(t,t')=\frac{\delta N_i(t)}{\delta h_i(t')}$ \textit{via} temporal integration:

\begin{eqnarray}
\fl \frac{\partial}{\partial t} \chi_i(t,t') = \chi_i(t,t') I_i'(N_i(t)) \left\lbrace \mu m(t) + \sigma \eta_i(t) \right\rbrace \nonumber
 \\
 + \chi_i(t,t') I_i'(N_i(t)) \gamma \sigma^2 \frac{p(p-1)}{2} \int_0^t \chi(t,s) C(t,s)^{p-2} J(N_i(s)) ds \nonumber\\
 + \chi_i(t,t') \left\lbrace R_i'(N_i(t)) + f_i'(N_i(t))\xi_i(t)  \right\rbrace \nonumber \\
 + I_i(N_i(t)) \gamma \sigma^2 \frac{p(p-1)}{2} \int_0^t \chi(t,s) C(t,s)^{p-2} J'(N_i(s)) \chi_i(s,t') ds \nonumber \\
 + I_i(N_i(t)) \delta(t-t')  \nonumber
\end{eqnarray}

We thus construct $\chi_i$ by temporal integration in $t$ at fixed $t'$, using the initial conditions $\chi_i(t,t')=0$ for $t<t'$ from causality. Eventually, we get:
$$\chi(t,t') = \mathbb{E} [ \chi_i(t, t') ] \sim \frac{1}{\#_{traj}} \sum_{i=1}^{\#_{traj}} \chi_i(t, t') $$

\section{Comparison of the different methods for the response function}
\label{si:responseSchemes}

\renewcommand{\arraystretch}{1.5}
\begin{center}

\begin{tabular}{|c|c|c|}
\hline 
\textbf{Method} & \textbf{Novikov} & \textbf{Temporal integration} \\ 
\hline 
\textbf{Formulation} & 
$\frac{1}{ \#_{traj}} \sum_{i=1}^{\#_{traj}} \frac{1}{\sigma} J ( N_i(t) ) \int ds C^{-1}_{t',s} \; \eta_i(s)$ & $\frac{1}{\#_{traj}} \sum_{i=1}^{\#_{traj}} \chi_i(t, t')$ \\ 
\hline 
\textbf{Needed $\#_{traj}$} & Non-linearity dependent & Low \\ 
\hline 
\textbf{Complexity} & $\mathcal{O}(\#_{traj} \; \#_{time}^2)$ & $\mathcal{O}(\#_{traj} \; \#_{time}^3)$ \\ 
\hline 
\textbf{Adequacy} & Linear problems & Non linear, but short range \\ 
\hline 
\end{tabular} 

\end{center}

\section{Closure in the Unique Equilibrium phase}

\subsection{Linear stability of dead species}
\label{siSec:stability_dead}
We consider the rLV system of equations \eref{eq:rLV}. For each species, there are two possible equilibria $0$ or $N_i^*=1-\sum_{j \neq i} \alpha_{ij} N_j$. In total, assuming the reduced matrix is almost always invertible (which is reasonable), this gives $2^S$ possible equilibria for the ecosystem, from which we would have to substract the unreachable ones with negative populations. We can linearize the equation around both possible choices for one species:
$$\dot{\delta N_i}= 
\cases{ N_i^* \delta N_i & if the fixed point is $0$,\\
-N_i^*\sum_{j \neq i} \alpha_{ij} \delta N_j & if the fixed point is $N_i^*$.
}$$
Eventually, we see that the dead species solution is linearly unstable if the alive solution $N_i^*$ is positive.
\subsection{Unique Equilibrium system of equations}
\label{siSec:1eq}
We now use the stationary cavity solution from equation \eref{eq:rsGuy}. The species population distribution is a truncated Gaussian: $p(n)=\phi p_+(n) + (1-\phi)\delta(n)$ where $\phi$ is the fraction of surviving species, and $p_+$ is a Gaussian distribution whose parameters need to be determined. We inject this form in the closure system \eref{eq:closEqRS}. We introduce the parameters $q=\mathbb{E}[N_\infty^2]$, $\Delta=(1-\mu m_\infty)\sigma^{-1}\sqrt{q}^{-1}$ and the functions $w_k(\Delta)=\int_{-\infty}^\Delta (\Delta-s)^k Ds$ where $Ds$ is the standard Gaussian measure, and eventually obtain:
\begin{equation}
\left\{
\eqalign{
		\frac{1-\sigma\sqrt{q}\Delta}{\mu} = \frac{\sigma \sqrt{q} }{1-\gamma \sigma^2 \chi_{int}} w_1(\Delta)\\
		\chi_{int} = \frac{1 }{1-\gamma \sigma^2 \chi_{int}} w_0(\Delta)\\
		1 = \frac{\sigma^2}{(1-\gamma \sigma^2 \chi_{int})^2} w_2(\Delta)
}
\right.
\end{equation}
It is worth noting that, with the implicit dependence $w_n=w_n(\Delta)$, the system can be rewritten as:
\begin{equation}
\label{eq:simplifSys}
\left\{
\eqalign{
	\sigma^2 \left( w_2 + \gamma w_0 \right)^2 = w_2\\
	1-\gamma\sigma^2 \chi_{int} = \sigma^2 \left( w_2 + \gamma w_0 \right)\\
	\sigma \sqrt{q}=\frac{\sigma^2 \left( w_2 + \gamma w_0 \right)}{\mu w_1 + \Delta \sigma^2 \left( w_2 + \gamma w_0 \right)}
}
\right.
\end{equation}
Under this form, the first line of system \eref{eq:simplifSys} gives $\Delta(\sigma,\gamma)$. Afterwards, we directly have $\chi_{int}(\sigma,\gamma,\Delta)$ and $q(\mu,\sigma,\gamma,\Delta)$.

\begin{figure}
\centering
\begin{minipage}{.5\textwidth}
  \centering
  \includegraphics[width=\linewidth]{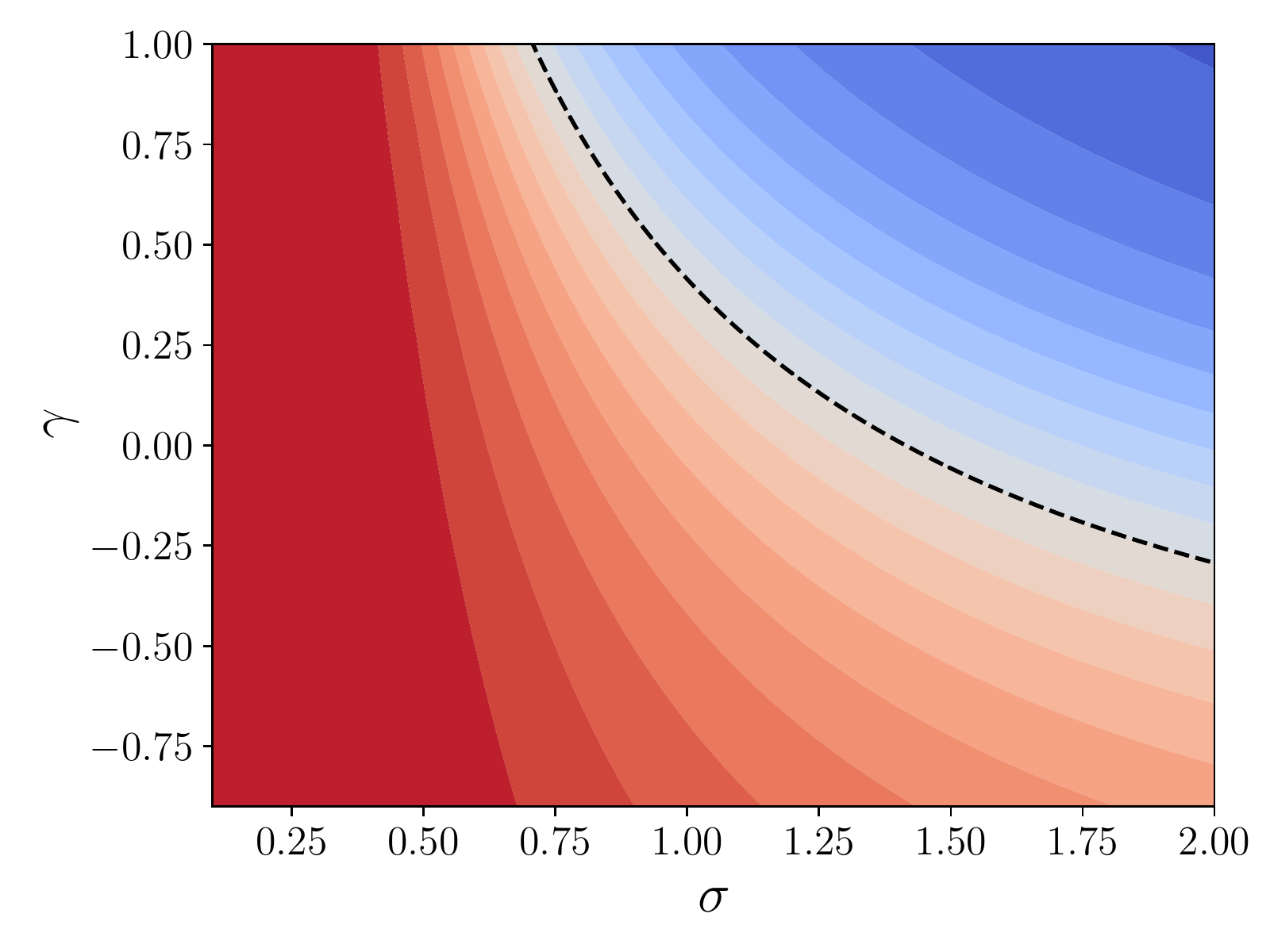}
\end{minipage}%
\begin{minipage}{.5\textwidth}
  \centering
  \includegraphics[width=\linewidth]{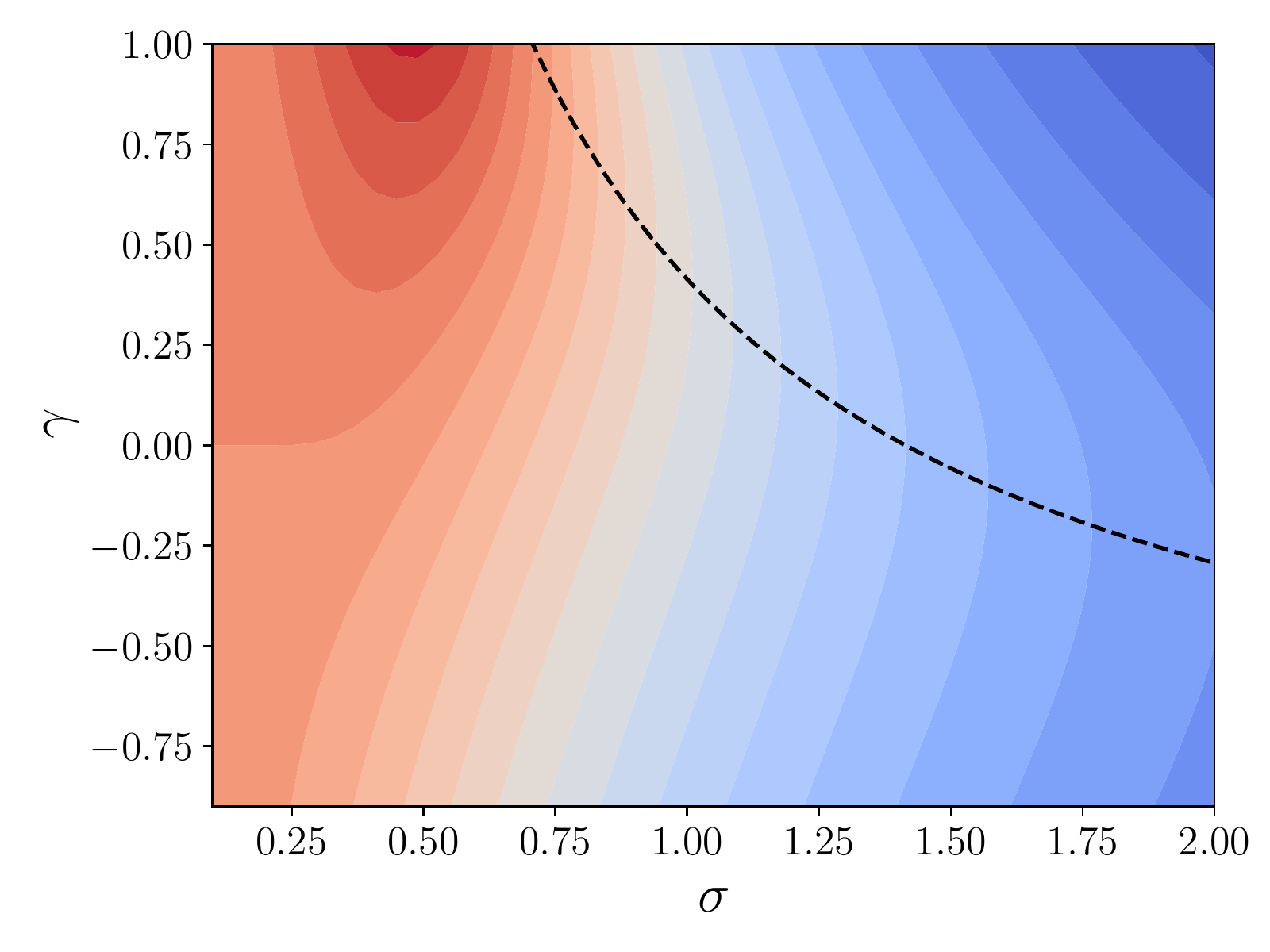}
\end{minipage}
\caption{Contour plots of the proportion of alive species $\phi$ and the integrated response kernel $\chi_{int}$, with parameters $\sigma$ and $\gamma$. Both of them are independent from $\mu$. In dotted black line is the chaotic transition. In the Multiple Attractors phase (upper right side of the black line), the stationary cavity analysis is only approximate. The color scale starts at $1$ in bright red, then each level corresponds to a $0.05$ decrease. 
Left: Proportion of alive species $\phi(\sigma,\gamma)$. It can be shown that the chaotic transition corresponds to an isocline $\phi=1/2$: half the species survive. Right: Integrated response kernel $\chi_{int}(\sigma,\gamma)$. It can be shown that the chaotic transition corresponds to a saddle line: $\partial_\gamma \chi_{int}|_{transition}=0$.}
\label{siFig:statCav}
\end{figure}

This system can be numerically solved in the variables $(\chi_{int},\Delta,q)$ as functions of the parameters $(\mu,\sigma,\gamma)$. All observables can then be computed from the solution. For example, the proportion of alive species $\phi=w_0(\Delta)$. On \fref{siFig:statCav}, we detail some analysis on $\phi(\sigma,\gamma)$, and the response to an environmental press $\chi_{int}(\sigma,\gamma)$.
See also \cite{buninPRE, galla_rLV, bunin_interaction_2016} for an analysis of the system.

\section{Linear stability analysis of the One-Equilibrium solution}
\label{si:stability_1eq}
Starting from the DMFT equation \eref{eq:DMFTrLV}, we linearize the system around a Unique Equilibrium. We introduce the relative amplitudes $\delta N(t) = N(t) - N_\infty$, $\delta m(t)=m(t)-m_\infty$ and $\delta \eta(t) = \eta(t) - \eta_\infty$, respectively corresponding to the population, average population and interaction noise. These amplitudes are supposed to go to zero, at least in the Unique Equilibrium phase. The self-consistent relation also holds for these relative amplitudes. Indeed, if we denote $C_c(t,s)$ the connected correlator, it verifies: 
$$C_c(t,s)=\mathbb{E}[\delta \eta(t)\delta \eta(s)]=\mathbb{E}[\delta N(t)\delta N(s)]$$

From now on, when we write the average $\mathbb{E}$, it will correspond to the average over both the static noise $\eta_\infty$ and the dynamical noise $\delta \eta$. The cases when $N_\infty=0$ will just give a relaxing exponential and not influence relevantly the correlator nor the response function at large times. Therefore, we will focus on the cases $N_\infty>0$, and write the corresponding average $\mathbb{E}_+$. The linearization reads:

\begin{equation}
\dot{\delta N}= - N_\infty( \delta N + \mu \delta m + \sigma \delta \eta  - \gamma \sigma^2 \int_0^t ds \chi(t,s) \delta N(s) - h)
\label{siEq:linCav}
\end{equation}

We focus on long times, and we assume time-translational invariance for the system: $C_c(t,s) = C(t-s)$ and $\chi(t,s)=\chi(t-s)$. This assumption has two consequences. First it transforms the integral term in equation \eref{siEq:linCav} into a convolution product. Secondly, denoting $\tilde{f}(\omega)$ the Fourier transform of $f$, the closure relations become:

\begin{equation}
\label{siEq:fourKC}
\fl 
\tilde{C}_{c,alive}(\omega)= \mathbb{E}_+\left[|\tilde{\delta N}(\omega)|^2\right]= \mathbb{E}_+\left[|\tilde{\delta \eta}(\omega)|^2\right] \qquad \tilde{\chi}_{alive}(\omega)=\mathbb{E}_+\left[\frac{\delta \tilde{N} (\omega)}{\delta \tilde{h}(\omega)}\right]
\end{equation}
We are computing observables $X$ for alive species only; the global average should be:
$$\mathbb{E}[X]=\phi \mathbb{E}_+[X] + (1-\phi)\mathbb{E}_{dead}[X]$$
where for relevant observables, the dead species contribution $\mathbb{E}_{dead}[X]$ vanishes in the large-$S$ limit. $\phi$ denotes the fraction of alive species in the Unique Equilibrium around which we are linearizing. It can be computed from the stationary cavity in \ref{siSec:1eq}. Now we send the linearized cavity equation \eref{siEq:linCav} into Fourier space:

\begin{equation}
\tilde{\delta N} = -(\mu \; \tilde{ \delta m } + \tilde{h} + \sigma \tilde{\delta \eta})\left(\frac{i \omega}{N_\infty} + 1 - \gamma \sigma ^2 \tilde{\chi}\right)^{-1}
\label{siEq:fourInt}
\end{equation}

Averaging directly equation \eref{siEq:fourInt}, and as the perturbation is of zero mean $\langle h \rangle_h=0$, we get that  the perturbation of the mean population is $\delta m = 0$. Finally we apply the relations \eref{siEq:fourKC}. The three terms $N_\infty$, $\tilde{h}$ and $ \tilde{\delta \eta}$ verify independence relations for different reasons:

\begin{itemize}
\item $\tilde{h}$ is independent of $N_\infty$ by construction, because we added the perturbation once the steady-state had already been reached;
\item $\delta \eta$ and $N_\infty$ are correlated in the temporal representation, but the Fourier correlation of stochastic variables only stands for the same pulsation. Therefore for any non-zero pulsation $\omega$, $\tilde{\delta \eta}(\omega)$ and $N_\infty$ are uncorrelated;
\item $h$ and $\delta \eta$ are directly uncorrelated, as the noise is sampled from a given covariance $C$.
\end{itemize}

$$|\frac{i \omega}{N_\infty} + 1 - \gamma \sigma ^2 \tilde{\chi}|^2|\tilde{\delta N}|^2 =  |\tilde{h}|^2 + \sigma^2 |\tilde{\delta \eta}|^2 +..$$

Eventually, we end up with the closed forms for the correlator and the response function:
\begin{eqnarray}
\tilde{C_c}(\omega) &= \left( \phi \mathbb{E}_+ \left[ |\frac{i \omega}{N_\infty} + 1 - \gamma \sigma^2 \tilde{\chi}(\omega)|^{-2}\right]^{-1} -\sigma^2 \right)^{-1} \label{siEq:Ctilde}\\
\tilde{\chi}(\omega) &= - \phi \mathbb{E}_+\left[ \left( \frac{i \omega}{N_\infty} + 1 - \gamma \sigma^2 \tilde{\chi}(\omega)\right)^{-1}\right] \label{siEq:Ktilde}
\end{eqnarray}
where the $\phi$ factor appears because we focus on the alive species. Indeed, in order to have normalization, the average $\mathbb{E}_+$ corresponds to averaging against the stationary measure which we rescaled: $1/\phi \; p_+(N_\infty)\; H(N_\infty) \; dN_\infty$. Therefore, the absolute contribution of alive species is $\phi\mathbb{E}_+$.

Now, as we are interested in the large time behaviour of the system, we perform a small $\omega$ expansion of the equations \eref{siEq:Ctilde} and \eref{siEq:Ktilde}. This limit needs to be taken carefully because there is a competitive effect in the average between $\omega$ and $\frac{1}{N_\infty}$. We finally get the expansions in equations \eref{eq:smallOmExpC} and \eref{eq:smallOmExpK}.

\section{Details of the numerical strategy for the DMFT solver}
\label{si:numStrategy}

We made a gitHub repository with the Python programs we wrote  \cite{felixroy_implements_2019}. There is also a runMe.py file which can be directly run in order to produce DMFT solutions and figures such as \fref{fig:convergenceObs} or \fref{fig:numPlateau}. In this section, we write down in details the methodology of the algorithm.

We discretize time in equal units of $dt$ such that $t_k = k \; dt$. We also fix the final time we're interested in as $t_{max} = \#_{time} \; dt$. We usually take $dt=0.1$. The two-dimensional functions then become matrices, and the one-dimensional ones are vectors:

$$m_k = m(t_k) \qquad  C_{kl} = C(t_k, t_l) \qquad \chi_{kl} = \chi(t_k, t_l) $$

We will now describe how one iteration of the algorithm is computed numerically. We start from the observables $m_k$, $C_{kl}$ and $\chi_{kl}$, and we want to compute the new ones $m^{new}_k$, $C^{new}_{kl}$ and $\chi^{new}_{kl}$ after one iteration.

\subsection{Sampling of the noise}

We will simulate $\#_{traj}$ trajectories that we will refer to as "species". Remember that they are independent in DMFT setting. We will then detail the procedure for one species only. For each species, we need a given realization of the gaussian noise at all times $\lbrace \eta_{k=1..\#_{time}} \rbrace = \lbrace \eta(t=0...t_{max}) \rbrace  $, sampled according to the correlator $C$. Given the discretization, we sample $\lbrace \eta_{k=1..\#_{time}} \rbrace$ as a multivariate gaussian vector with covariance $C_{kl}$. One way to do this is to diagonalize the matrix $C_{kl}$, then in the proper basis all components are independent.

\subsection{Numerical integration of the trajectory}

For the trajectory of the species, the integration of the differential equation is done with a basic Euler scheme. The Lotka-Volterra system is better simulated in log space. Therefore, if we denote $N_k = N(t_k)$, we implement the scheme:

$$\log N_{k+1} = \log N_k + dt \; \mathcal{F}(  N_k | m, \eta, \chi) + dt \; \mathcal{G}(  N_k | \lambda) $$

with:

\begin{equation*}
\left\{
\eqalign{
	&\mathcal{F}(  N_k | m, \eta, \chi) = 1 - N_k - \mu m_k - \sigma \eta_k + \gamma \sigma^2 dt \sum_{l=0}^k \chi_{kl} \; N_l\\
	&\mathcal{G}(  N_k | \lambda) = \cases{ 0&for $\lambda=0$\\
\exp \left[ \log(\lambda) - \log N_k \right]  &for $\lambda>0$\\}\\
}
\right.
\end{equation*}

The last $\lambda$-dependent scheme is for numerical stability whenever there is immigration in the system. Using this scheme, we compute the trajectory $N_{k=1..\#_{time}}$.

\subsection{Computing the new observables from the trajectories}

We sample and integrate the trajectories for $\#_{traj}$ species following the previous procedure. We end up with an array of trajectories:

$$\left\lbrace N_{k=1..\#_{time}}^{i=1..\#_{traj}} \right\rbrace$$

From them we can compute the new observables $m$ and $C$ by direct averages:

\begin{equation*}
\left\{
\eqalign{
	m^{new}_k&=  \frac{1}{\#_{traj}} \sum_{i=1}^{\#_{traj}} N_k^i\\
	C^{new}_{kl}&=\frac{1}{\#_{traj}} \sum_{i=1}^{\#_{traj}} N_k^i N_l^i\\
}
\right.
\end{equation*}

As stated in \ref{si:responseSchemes} and section \ref{numSolve}, the response function $\chi$ is more difficult to compute. The two methods can be used:

\begin{equation*}
\chi^{new}_{kl}=\cases{ 
	\frac{1}{\#_{traj}} \sum_{i=1}^{\#_{traj}} \frac{1}{\sigma} N_k^i \; dt \; \sum_{l'=1}^{\#_{time}} C^{-1}_{ll'} \eta_{l'}^i&for Novikov\\
	\frac{1}{\#_{traj}} \sum_{i=1}^{\#_{traj}} \chi_{kl}^i & for temporal integration\\}
\end{equation*}

In the last line of the equation $\chi_{kl}^i$ is integrated for each species according to \ref{si:tempIntegResponse}. Eventually, we will start a new iteration of the algorithm, with a soft update:

\begin{equation*}
\left\{
\eqalign{
	m^{updated}&=  (1-a)\; m + a\; m^{new}\\
	C^{updated}&=(1-a)\; C + a\; C^{new}\\
	\chi^{updated} &= (1-a)\; \chi + a\; \chi^{new}
}
\right.
\end{equation*}

After some trials, a reinjection parameter $a=0.3$ is a good value.

\subsection{Convergence and the iterative strategy}

After one iteration of the algorithm, the new results always present some statistical noise, due to the fact that we average over a finite number $\#_{traj}$ of trajectories. To get a good convergence, we increase the number of trajectories as the iteration goes on. The first iterations are performed with few trajectories; they correspond to rough steps in the configurational space. As the observables get closer to the real solution, we refine the iterations by using more trajectories.

\begin{figure}[hbtp]
\centering
\includegraphics[width=0.6\linewidth]{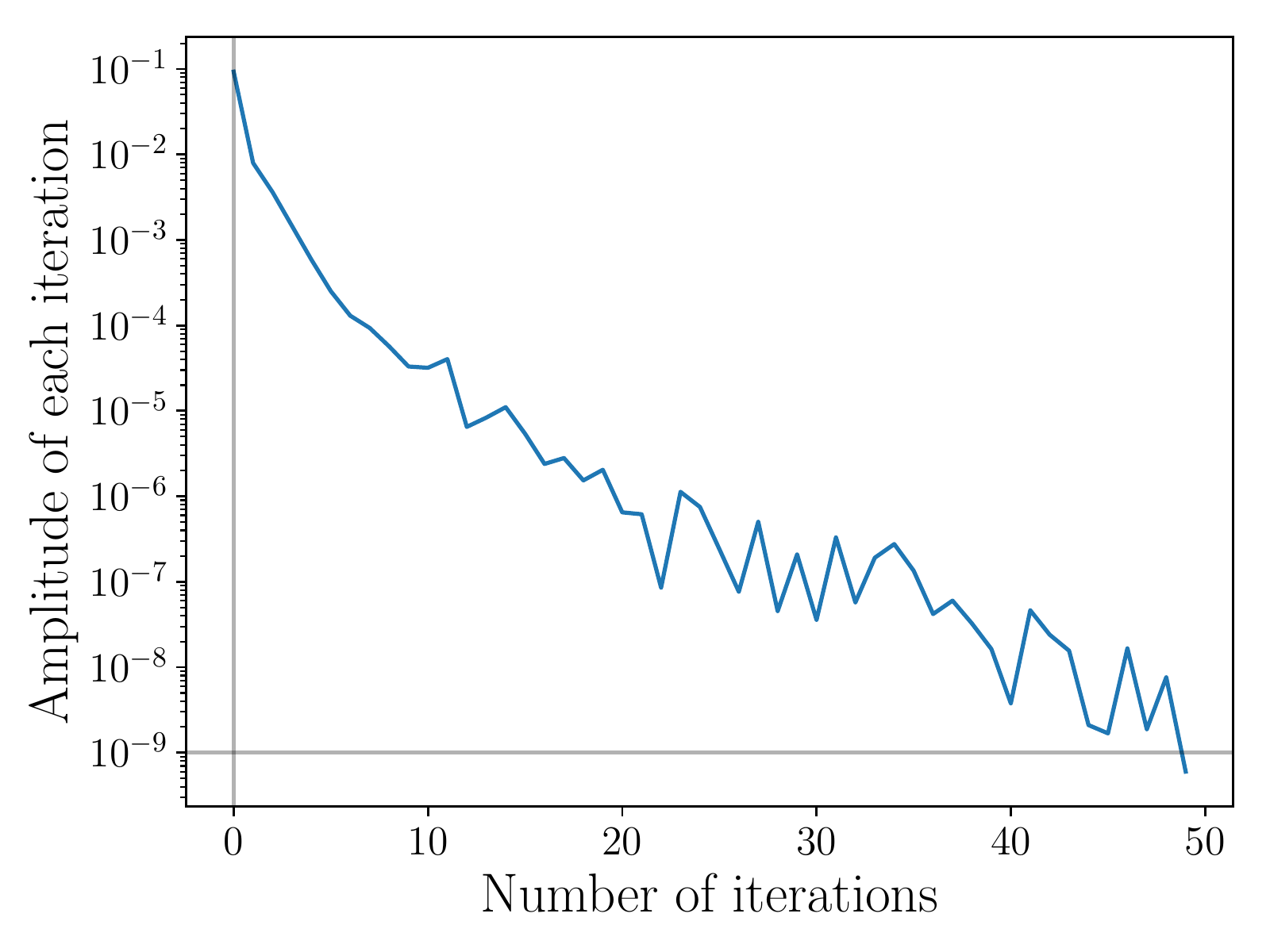}
\caption{Amplitude of each iteration step as a function of the number of iterations. The amplitude is computed as the matrix norm of the difference in the correlator before and after iteration. It is plotted on a semilog scale. This computation was done with parameters $(\mu,\sigma,\gamma,\lambda)=(10,4,-1,10^{-4})$ in the Unique Equilibrium phase.}
\label{siFig:iteNorm}
\end{figure}

All results are shown with the following scheme: $30$ iterations with $10^3$ trajectories each, then $10$ iterations with $10^4$ each, and $20$ iterations with $10^5$ each. The convergence is considered to be reached when the iteration step becomes lower than a given threshold. More precisely, labeling $C^i_{kl}$ the correlator after iteration $i$, we have reached convergence when:
$$\|C^{i+1}-C^i\|_F < 10^{-9}$$
where $\|M\|_F=\#_{time}^{-2}\sum_{kl}M_{kls}^2$ is the rescaled Frobenius norm. We use the threshold on the correlator, because we found that it is the most difficult observable to converge. A mixed criterion in all three observables would work as well. On \fref{siFig:iteNorm} we show the convergence in terms of iteration steps.

\section{Some examples of numerical solutions}
\label{si:numExamples}

\begin{figure}[hbtp]
\centering
\includegraphics[scale=0.7]{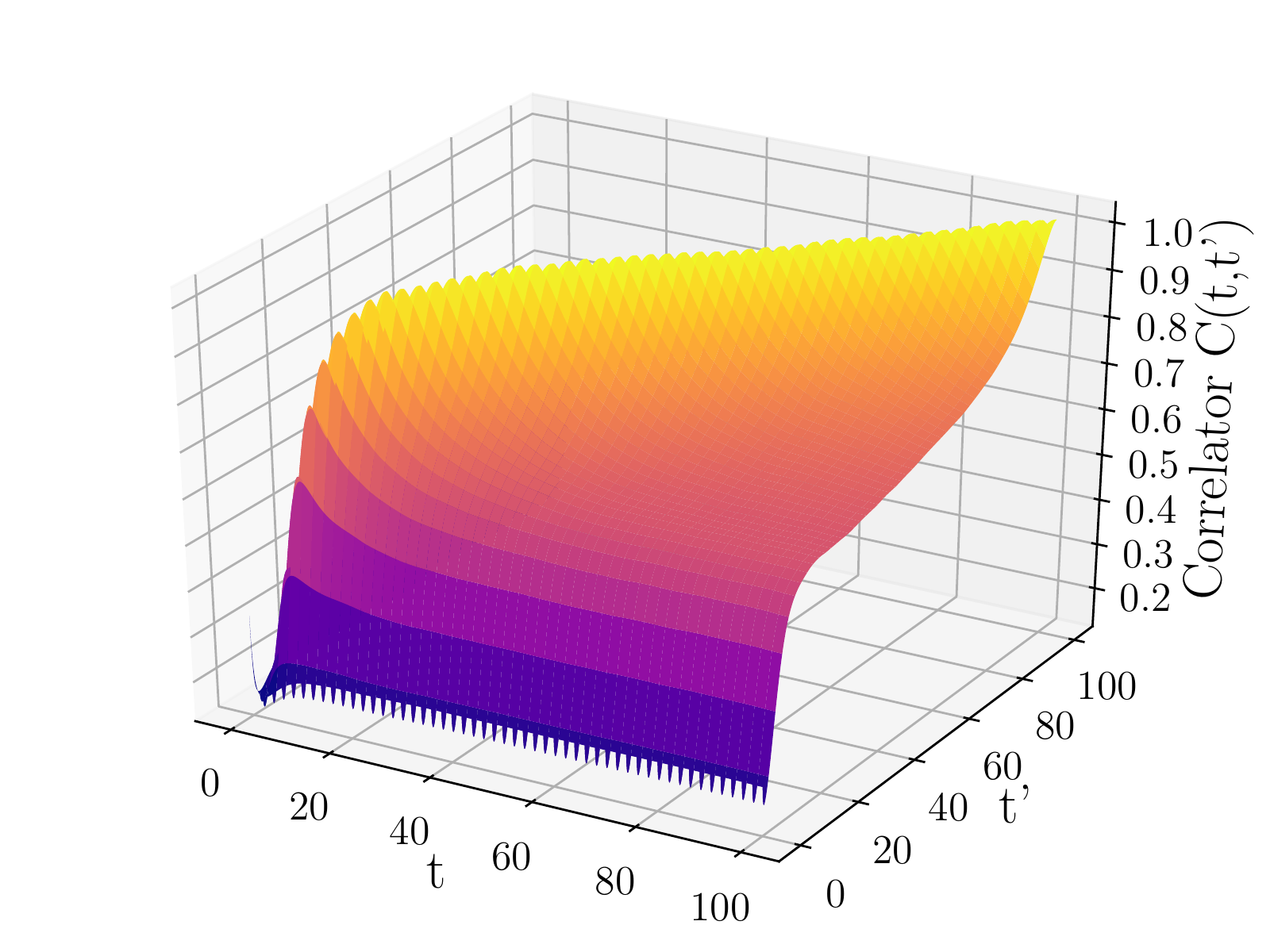}
\caption{Numerical correlator, for rLV DMFT with parameters $(\mu,\sigma,\gamma,\lambda)=(4,2,0,10^{-4})$ in the Multiple Attractors phase. The parameters of the program are the same as in \fref{fig:convergenceObs}. Contrary to the Unique Equilibrium case in \fref{fig:numPlateau}, there is no convergence towards a plateau. However, after a transient, the systems becomes TTI.}
\label{siFig:numPlateau}
\end{figure}

\begin{figure}[hbtp]
\centering
\includegraphics[scale=0.8]{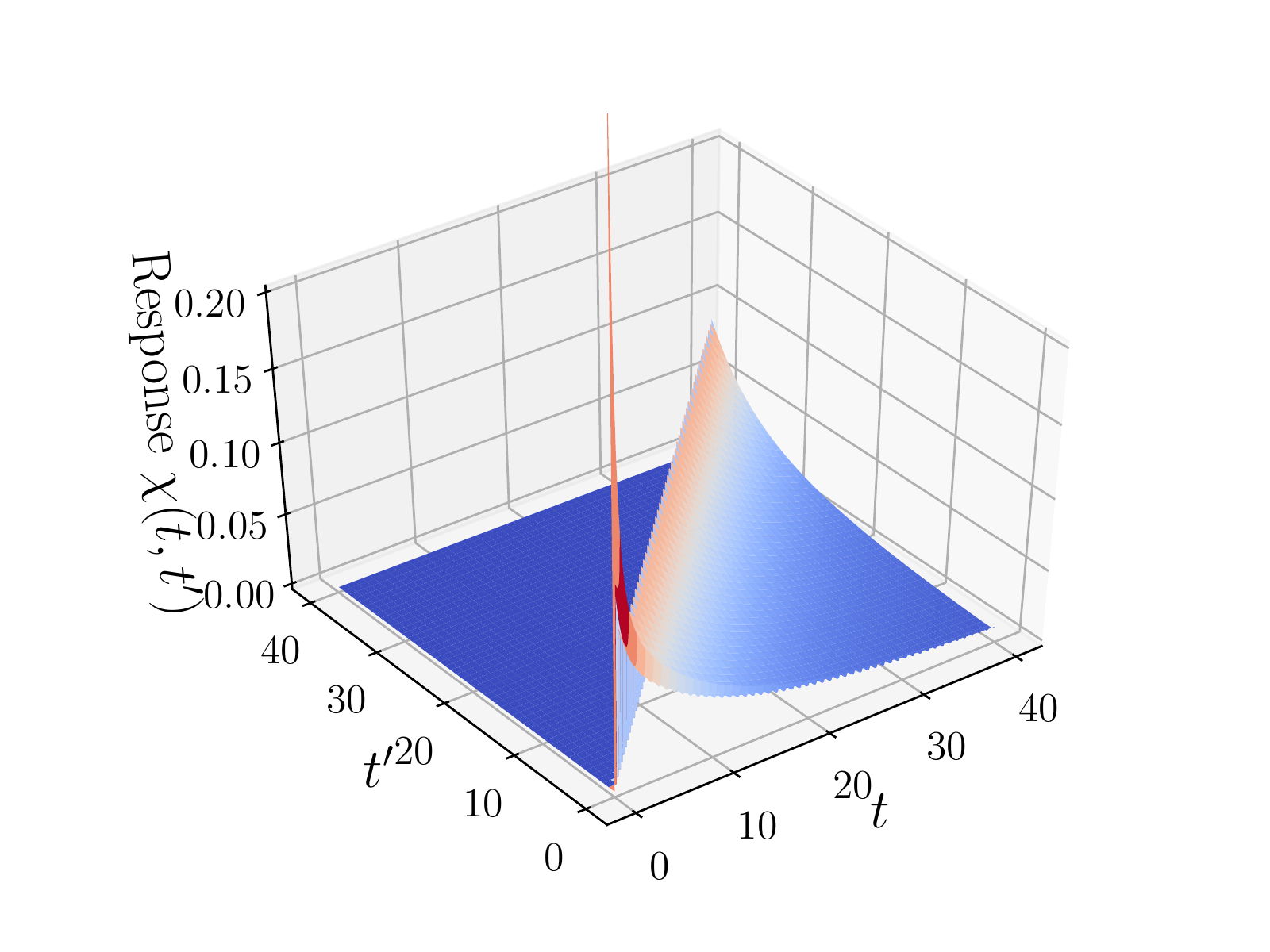}
\caption{Numerical response function, for rLV DMFT with parameters $(\mu,\sigma,\gamma,\lambda)$ $=(10,1/2,1/2,10^{-4})$ below the onset of chaos. The parameters of the program are the same as in \fref{fig:convergenceObs}. From causality, $\chi(t,t')=0$ for $t<t'$. It can be shown analytically that $\chi(t,t)=m(t)$. Then, for $t>t'$, there is a relaxation towards 0, as the perturbation is absorbed.}
\label{siFig:numResponse}
\end{figure}

On \fref{siFig:numPlateau}, we show an example of a chaotic correlator $C(t,t')$. It is to be put in contrast with \fref{fig:numPlateau}, which depicted the Unique Equilibrium plateau type correlator. On \fref{siFig:numResponse}, we show an example of a response numerical solution $\chi(t,t')$. The behaviour of $\chi$ does not seem to change drastically between Unique Equilibrium phase and the Multiple Attractors phase.

\clearpage


\bibliographystyle{unsrt}      
\bibliography{biblioEcosystems.bib}   

\end{document}